\documentstyle[epsf,12pt]{article}
\makeatletter

 \@addtoreset{equation}{section}
\makeatother

\begin{document}
\topmargin -35pt
\oddsidemargin 5mm

\newcommand {\beq}{\begin{eqnarray}}
\newcommand {\eeq}{\end{eqnarray}}
\newcommand {\non}{\nonumber\\}
\newcommand {\eq}[1]{\label {eq.#1}}
\newcommand {\defeq}{\stackrel{\rm def}{=}}
\newcommand {\gto}{\stackrel{g}{\to}}
\newcommand {\hto}{\stackrel{h}{\to}}
\newcommand {\1}[1]{\frac{1}{#1}}
\newcommand {\2}[1]{\frac{i}{#1}}
\newcommand {\th}{\theta}
\newcommand {\thb}{\bar{\theta}}
\newcommand {\ps}{\psi}
\newcommand {\psb}{\bar{\psi}}
\newcommand {\ph}{\varphi}
\newcommand {\phs}[1]{\varphi^{*#1}}
\newcommand {\sig}{\sigma}
\newcommand {\sigb}{\bar{\sigma}}
\newcommand {\Ph}{\Phi}
\newcommand {\Phd}{\Phi^{\dagger}}
\newcommand {\Sig}{\Sigma}
\newcommand {\Phm}{{\mit\Phi}}
\newcommand {\eps}{\varepsilon}
\newcommand {\del}{\partial}
\newcommand {\dagg}{^{\dagger}}
\newcommand {\pri}{^{\prime}}
\newcommand {\prip}{^{\prime\prime}}
\newcommand {\pripp}{^{\prime\prime\prime}}
\newcommand {\prippp}{^{\prime\prime\prime\prime}}
\newcommand {\delb}{\bar{\partial}}
\newcommand {\zb}{\bar{z}}
\newcommand {\mub}{\bar{\mu}}
\newcommand {\nub}{\bar{\nu}}
\newcommand {\lam}{\lambda}
\newcommand {\lamb}{\bar{\lambda}}
\newcommand {\kap}{\kappa}
\newcommand {\kapb}{\bar{\kappa}}
\newcommand {\xib}{\bar{\xi}}
\newcommand {\Ga}{\Gamma}
\newcommand {\rhob}{\bar{\rho}}
\newcommand {\etab}{\bar{\eta}}
\newcommand {\tht}{\tilde{\th}}
\newcommand {\zbasis}[1]{\del/\del z^{#1}}
\newcommand {\zbbasis}[1]{\del/\del \bar{z}^{#1}}
\newcommand {\vecv}{\vec{v}^{\, \prime}}
\newcommand {\vecvd}{\vec{v}^{\, \prime \dagger}}
\newcommand {\vecvs}{\vec{v}^{\, \prime *}}
\newcommand {\alpht}{\tilde{\alpha}}
\newcommand {\xipd}{\xi^{\prime\dagger}}
\newcommand {\pris}{^{\prime *}}
\newcommand {\prid}{^{\prime \dagger}}
\newcommand {\Jto}{\stackrel{J}{\to}}
\newcommand {\vprid}{v^{\prime 2}}
\newcommand {\vpriq}{v^{\prime 4}}
\newcommand{\vs}[1]{\vspace{#1 mm}}
\newcommand{\hs}[1]{\hspace{#1 mm}}

\begin{titlepage}

\begin{flushright}
OU-HET 350\\
TIT/HEP-451\\
hep-th/0006038\\
June 2000
\end{flushright}
\bigskip

\begin{center}
{\LARGE\bf
Geometry and the Low-Energy Theorem\\}

{\LARGE\bf
in $N=1$ Supersymmetric Theories
}
\vs{10}

\bigskip
{\renewcommand{\thefootnote}{\fnsymbol{footnote}}
{\large\bf Kiyoshi Higashijima$^a$\footnote{
     E-mail: higashij@phys.sci.osaka-u.ac.jp.}
 and Muneto Nitta$^b$\footnote{
E-mail: nitta@th.phys.titech.ac.jp}
}}

\setcounter{footnote}{0}
\bigskip

{\small \it
$^a$Department of Physics,
Graduate School of Science, Osaka University,\\
Toyonaka, Osaka 560-0043, Japan\\
$^b$Department of Physics, Tokyo Institute of Technology, 
Oh-okayama, \\ Meguro, Tokyo 152-8551, Japan\\
}
\end{center}
\bigskip
 
\begin{abstract}
We investigate geometrical structures and  
low-energy theorems  
of $N=1$ supersymmetric nonlinear sigma models in four dimensions.
When a global symmetry spontaneously breaks down to its subgroup, 
the low-energy effective Lagrangian of massless particles 
is described by a supersymmetric nonlinear sigma model
whose target manifold is parametrized by 
Nambu-Goldstone (NG) bosons and quasi-NG (QNG) bosons.
The unbroken symmetry changes 
at each point in the target manifold 
and some QNG bosons change to NG bosons 
when unbroken symmetry become smaller.
The QNG-NG change and their interpretation is shown 
in a simple example, the $O(N)$ model.
We investigate low-energy theorems at general points.
\end{abstract}

\end{titlepage}

\section{Introduction}
In non-supersymmetric theories, the Nambu-Goldstone (NG) theorem 
tells us that there appear as many massless NG bosons as the 
number of broken generators,  namely ${\rm dim}\; (G/H)$, 
when a global symmetry $G$ spontaneously breaks down to 
its subgroup $H$. The NG bosons parameterize a vacuum 
degeneracy which has one-to-one correspondence with the freedom 
of the embedding $H$ into $G$. 
The effective Lagrangian of massless bosons can be expanded by the 
number of space-time derivatives, and the leading term, 
with two derivatives, is described by nonlinear sigma models 
on target manifolds $G/H$,  
whose coordinates are parametrized by NG bosons. On this manifold, 
the unbroken symmetry $H$ is realized linearly, while the broken 
symmetry $G$ is realized nonlinearly by NG bosons~\cite{CCWZ}.
For non-supersymmetric cases, low-energy theorems tell us that 
low-energy scattering amplitudes of NG bosons 
are determined solely by the symmetries $G$ and $H$, 
and do not depend on details of the underlying 
theory (for a review, see Ref.~\cite{BKY}).
The effective Lagrangian reproduces these low-energy theorems. 

In supersymmetric theories, there appear additional massless bosons 
called quasi-NG (QNG) bosons~\cite{KOY} 
(and their fermionic superpartners).\footnote{
Fermions, called the QNG fermions, would be interesting particles, 
when we regard quarks and leptons as QNG fermions.
But we do not discuss QNG fermions in this paper.
} 
Leading terms of massless effective Lagrangian 
are described by $N=1$ supersymmetric nonlinear sigma models 
(for example see Ref.~\cite{CL1}). 
Target manifolds of $N=1$ nonlinear sigma models are 
K\"{a}hler manifolds~\cite{Zu}:  
A manifold whose metric is given by a K\"ahler potential 
$K(\ph,\ph^*)$
$$g_{ij^*}(\ph,\ph^*)
={\del^2K(\ph,\ph^*)\over \del \ph^i \del \ph^{*j}},$$
is called a K\"ahler manifold. $\ph (x)$ is a complex scalar 
component of a chiral superfield. 
NG and QNG bosons are coordinates of a complex coset 
manifold $G^{\bf C}/\hat H$, where $G^{\bf C}$ is 
the complexification of $G$ and $\hat H$ is the 
complex subgroup often larger than $H^{\bf C}$, 
the complexification of $H$. K\"{a}hler potentials of 
$G^{\bf C}/\hat H$ have been constructed by Bando, 
Kuramoto, Maskawa and Uehara (BKMU)~\cite{BKMU}
(for a review, see Ref.~\cite{Ku}).
If $\hat H = H^{\bf C}$, the number of QNG bosons is 
the same as that of NG bosons, 
and nonlinear realizations in these cases are called  
``maximal realizations'' or ``fully-doubled realizations''.
On the other hand, if $\hat H (\supset H^{\bf C})$ becomes 
larger, the number of QNG bosons decreases. If there is no 
QNG boson, realizations are called ``pure realizations'',  
and studied extensively~\cite{IKK}. Pure realizations cannot 
be obtained as the low-energy limit of underlying linear 
theories since there remains at least 
one QNG boson~\cite{Le,BL,KS}.
If there is, however, gauge symmetry, 
it is possible to absorb pairs 
of a QNG and a NG bosons by the supersymmetric Higgs mechanism. 
Hence pure realizations are in some cases obtained as 
low-energy theories of gauged linear sigma models~\cite{Ku,HN}.

To investigate low-energy theories with supersymmetry,
it is important to understand geometric structures of  
supersymmetric sigma models. In the cases of pure realizations, 
the geometry of the target space is well understood, because 
K\"ahler potentials are uniquely determined by the metric 
of $G/H$. When there are QNG bosons, however, the coset space, 
$G/H$ where NG bosons reside, is a subspace of the target space. 
Since the metric in directions along QNG bosons is not 
determined by the geometry of its subspace $G/H$, 
the effective Lagrangian is not unique in this case 
and depends on an arbitrary function of many 
$G$-invariant variables~\cite{KS,Ni}. 
When there are many $G$-invariant variables, 
it is complicated to study geometric structures 
of target spaces in general. 
In this paper, we investigate 
the $O(N)$ model whose K\"{a}hler potential contains an 
arbitrary function of a single variable, 
but generalizations to other models are straightforward. 

\bigskip
This paper is organized as follows.
We review our previous results~\cite{HNOO}
in the rest of this section.
The low-energy theorems at the symmetric points are explained.
In Sect.~2, we study non-symmetric points 
where unbroken symmetry $H$ is reduced to a smaller group $H\pri$.
In Sect.~3, to investigate the geometrical structure of 
the supersymmetric nonlinear sigma model with $O(N)$ symmetry, 
we show explicitly how the different compact homogeneous 
manifolds $G/H$ and $G/H\pri$ are embedded in the full target 
manifold $G^{\bf C}/\hat H$ by using the method of Shore~\cite{Sh}.
We see how some QNG bosons change to NG bosons at the 
non-symmetric points.
In Sect.~4, we derive the low-energy theorems of 
NG and QNG bosons at the general points 
of the target manifold when the K\"{a}hler potential 
is the simplest one.  
Sect.~5 is devoted to conclusion and discussion.  
In Appendix A, we explain the K\"{a}hler normal coordinate 
which is used to calculate the low-energy theorems.
In Appendix B, 
some geometric quantities are calculated for 
the most general $O(N)$-invariant model.

\bigskip
The general low-energy effective Lagrangian 
of massless bosons $\phi^{\alpha}(x)$
is a nonlinear sigma model whose target manifold  
has the metric $g_{\alpha\beta}(\phi)$, 
\beq
 {\cal L} =\1{2} g_{\alpha\beta}(\phi)
  \del_{\mu}\phi^{\alpha} \del^{\mu}\phi^{\beta} .
\eeq
Low-energy scattering amplitudes are 
unchanged by a field redefinition, 
which is a general coordinate transformation 
in the target manifold. 
By expanding this in the Riemann normal coordinate $\phi^i$~\cite{AFM}  
up to the forth order, 
and regarding the fourth order terms 
as interaction terms ${\cal L}_{\rm int}$, 
low-energy two-body scattering amplitudes 
of the massless bosons $\phi^i$ (with momenta $p_i$) 
\beq
&& \left< \phi^k(p_k),\phi^l(p_l)|i {\cal L}_{int} 
    |\phi^i(p_i),\phi^j(p_j) \right> \non
& &= i(2\pi)^4 \delta^{(4)}(p_k+p_l-p_i-p_j)
   {\cal M}(\phi^i(p_i),\phi^j(p_j) \to \phi^k(p_k),\phi^l(p_l)) 
\eeq
can be calculated by summing up 
all the tree graphs\footnote{
To calculate the next-to leading order ${\cal O}(p^4)$,
we need to sum up one-loop graphs of the leading term and 
tree graphs of the four derivative terms, 
with obeying Weinberg's counting theorem.}, given by   
\beq
 {\cal M}(\phi^i, \phi^j \to \phi^k, \phi^l) 
 = -\1{3f_{\pi}^4}[(s-u)R_{kijl} + (u-t)R_{ijkl} + (t-s)R_{kjli}] .
 \label{scat.amp.}
\eeq
Here $f_{\pi}$ is the decay constant of the NG bosons (pions),  
$R_{ijkl}$ is the curvature tensor of the target manifold, 
and we have defined the Mandelstam's variables by  
\beq
 && s \defeq (p_i + p_j)^2 = +2 p_i \cdot p_j = +2 p_l \cdot p_k,  \non
 && t \defeq (p_i - p_k)^2 = -2 p_i \cdot p_k = -2 p_l \cdot p_j,  \non 
 && u \defeq (p_i - p_l)^2 = -2 p_i \cdot p_l = -2 p_j \cdot p_l.
\eeq
We consider cases that a global symmetry $G$ spontaneously 
breaks down to its subgroup $H$.
We express broken and unbroken generators by~\footnote{
The Lie algebras of the groups $G$ and $H$ are denoted
by ${\cal G}$ and ${\cal H}$, respectively.
}
\beq
 X_i \in {\cal G} - {\cal H}, \hs{10}
 H_a \in {\cal H}, \hs{10}
 (T_A \in {\cal G}).
\eeq
In the cases of symmetric spaces $G/H$  
(in which there is a symmetry $X_i \to - X_i\;,\; H_a \to H_a$),  
the curvature tensor can be calculated 
by using structure constants of $G$, ${f_{AB}}^C$, to yield
\beq
 R_{ijkl} = f_{\pi}^2 \; {f_{ij}}^a f_{akl}, 
 \label{curv.of irre.sym.}
\eeq
and the low-energy theorems become
\beq
 {\cal M}(\phi^i, \phi^j \to \phi^k, \phi^l)
 = -\1{3f_{\pi}^2}[(s-u){f_{ki}}^a f_{ajl} + (u-t){f_{kl}}^a f_{aij}
                   + (t-s){f_{kj}}^a f_{ali} ].
     \label{LET non-SUSY}
\eeq

In $N=1$ supersymmetric theories,
the low-energy effective Lagrangian of 
massless chiral superfields 
$\Ph^i(x,\th,\thb) = \ph^i(x) 
+ \sqrt{2}\th\ps^i(x) + \th\th F^i(x)$ 
(where $\ph^i$, $\ps^i$ and $F^i$ are complex scalar fields, 
Weyl fermions, auxiliary scalar fields, respectively.)
is a supersymmetric nonlinear sigma model~\cite{WB},  
\beq
 {\cal L} &=& \int d^2\th d^2\thb K(\Ph,\Phd) \non
 &=& g_{ij^*}(\ph,\ph^*)\del_{\mu}\ph^i \del^{\mu}\ph^{*j}
 + ig_{ij^*}\psb^j \sigb^{\mu}
   (\del_{\mu}\ps^i + {\Gamma^i}_{lk}\del_{\mu}\ph^l \ps^k) \non
 &&+ \1{4} R_{ij^*kl^*} \ps^i\ps^k \psb^j\psb^l . \label{SNLSM}
\eeq
Here the metric tensor is calculated by the K\"{a}hler potential as
\beq
 g_{ij^*}(\ph,\ph^*)
 = \del_i \del_{j^*} K(\ph,\ph^*) , 
\eeq
and $R_{ij^*kl^*}$ and ${\Gamma^i}_{lk}$ are 
the complex curvature 
and the connection, respectively.
In Eq.~(\ref{SNLSM}), the auxiliary fields $F^i$ have been eliminated 
by using their equations of motion.
The massless chiral NG superfields appear when the global symmetry $G$ 
spontaneously breaks down to its subgroup $H$ with preserving $N=1$ 
supersymmetry. 
We denote complex broken and unbroken generators as~\footnote{
Complex generators are complex linear combinations of 
Hermitian generators of ${\cal G}$. 
We use indices $R,S,T$ and $L,M,N$ for 
{\it complex} broken and unbroken generators, respectively.
}
\beq
 Z_R \in {\cal G}^{\bf C} - \hat{\cal H} ,\hs{10} 
 K_M \in \hat{\cal H} .
\eeq
Their commutation relations are
\beq
 \left[ K_M, K_N \right] = i {f_{MN}}^L K_L,\hs{5}
 \left[ Z_R, K_M \right] = i {f_{RM}}^S Z_S,\hs{5}
 \left[ Z_R, Z_S \right] = i {f_{RS}}^M K_M ,
\eeq
where we have assumed the existence of an automorphism
\beq
 Z_R \to - Z_R ,\hs{10} 
 K_M \to K_M .  \label{auto mor.}
\eeq
The target manifold 
(which is a $G^{\bf C}$-orbit of vacuum vector $\vec{v}$) 
is a complex coset manifold $G^{\bf C}/\hat H$, 
and its representative is
\beq
 \xi(\Ph) = {\rm e}^{i \Ph \cdot Z} 
     \in G^{\bf C}/\hat H,\hs{10} 
 \Phi \cdot Z = \sum_{i=1}^{N_{\Phi}} \Phi^i Z_R \delta^R_i .
\label{bkmu}
\eeq
Here $\Ph^i(x,\th,\thb)$ are the NG chiral superfields 
and $N_{\Phi}$ is a number of $\Phi^i$. 
The left action of $G$ on the coset representative is
\beq
 \xi \stackrel{g}{\to} \xi ^{\prime} = g \xi \hat h^{-1}(g,\xi), 
  \hs{10} g\in G,  \label{left-act.}
\eeq
where $\hat h(g,\xi)\in \hat H$ is called an $\hat H$-compensator.

It is known that, when a vacuum vector $\vec{v}$ 
is in the real representation of $G$, 
or when $G/H$ is a symmetric space,  
only maximal realizations are possible~\cite{Le}.
So we discuss maximal realizations, 
where there appear the same numbers of NG and QNG bosons 
(at a symmetric point defined below). 
The low-energy effective K\"{a}hler potential 
can be written as~\cite{Le,BKMU,Sh,KS}
\beq
 K(\Ph,\Phd)
  = f(\vec{v}\,\dagg \xi^{\dagger}(\Ph \dagg)\xi(\Ph) \vec{v}) \;, 
  \label{A type}
\eeq
where $f$ is an {\it arbitrary} function, 
which cannot be determined by symmetry.\footnote{
If there are some $G$-invariants, a K\"{a}hler potential 
can be written as an arbitrary function of such invariants.
} 
This arbitrariness is a characteristic feature of 
non-pure realizations. 
Note that this K\"{a}hler potential 
is $G$-invariant by Eq.~(\ref{left-act.})
but $not$ $G^{\bf C}$-invariant: 
A $G$-action is a general coordinate transformation 
preserving the metric (the K\"{a}hler potential),
while a $G^{\bf C}$-action does not preserve the metric. 
This fact has an important consequence: The symmetry of the 
action is still the compact real group $G$, although the target 
space is a $G^{\bf C}$-orbit of vacuum vector $\vec{v}$.

A holomorphic vielbein $E^R_i$ and 
a canonical $\hat H$-connection $W^M_i$ can  
be read as coefficients of broken and unbroken elements 
of the Maurer-Cartan $1$-form
\beq
 \1{i}\xi(\ph)^{-1} d\xi(\ph)
 = ( E^R_i(\ph) Z_R + W^M_i(\ph) K_M ) d\ph^i . \label{MC1form2}
\eeq
We define {\it symmetric points} by 
points with the largest unbroken symmetry. 
(As seen in the next section, 
there exist points with smaller unbroken symmetry.) 
We can take a coordinate system $\ph$ on 
$G^{\bf C}/\hat H$ so that 
the origin of $\ph$ is a symmetric point. 
Then at a symmetric point $\ph=0$, 
the vielbein and the $\hat H$ connection take 
the form of 
\beq
 E^R_i|_{\ph =0} = {\delta}^R_i ,\hs{10} 
 W^M_i|_{\ph =0} = 0, 
\eeq
respectively, 
and differentiations of the vielbein with respect to coordinates  
at the point are
\beq
 \del_j E^R_i|_{\ph =0} = 0  \;.\label{der.E}
\eeq
\ From Eqs.~(\ref{MC1form2}) to (\ref{der.E}), 
We can calculate the curvature tensor at the symmetric point, 
given by
\beq
  R_{ij^*kl^*}
  = f_1 (\vec{v}\dagg Z_S\dagg Z_V\dagg Z_U Z_R \vec{v})
    \delta^R_i (\delta^S_j)^* \delta^U_k (\delta^V_l)^*
   + g^2 (\delta_{ik}\delta_{j^*l^*} + \delta_{ij^*}\delta_{kl^*}
           +\delta_{il^*}\delta_{kj^*}) ,\hs{5} 
 \label{curve.Kahler}
\eeq
where we have defined $v^2 \defeq \vec{v}\dagg \vec{v}$,  
$f_1 \defeq f' (v^2),\; f_2 \defeq f'' (v^2)$ etc.    
and a constant $g$ by
\beq
 g^2 \defeq f_2 \; v^2  .
\eeq
We express complex scalar fields by $\ph^i(x) = A^i(x) + i B^i(x)$, 
where $A^i(x)$ and $B^i(x)$ are real scalar fields. 
In maximal realization cases, 
$A^i(x)$ and $B^i(x)$ are NG and QNG bosons, respectively. 
In the real basis of $A^i$ and $B^i$, the K\"{a}hler condition on 
the curvature tensor becomes  
\beq
 \cases{
     R_{A^iA^jA^kA^l} = R_{B^iB^jB^kB^l}
  =  R_{A^iA^jB^kB^l} = R_{B^iB^jA^kA^l}, \cr
     R_{B^iA^jB^kA^l} = R_{A^iB^jA^kB^l}
  =- R_{B^iA^jA^kB^l} = -R_{A^iB^jB^kA^l}, \cr
     R_{B^iA^jA^kA^l} = - R_{A^iB^jA^kA^l}
 = - R_{A^iB^jB^kB^l} = R_{B^iA^jB^kB^l}, \cr
     R_{A^iA^jB^kA^l} = - R_{A^iA^jA^kB^l}
 = - R_{B^iB^jA^kB^l} = R_{B^iB^jB^kA^l} .\cr} \label{sym.of R}
\eeq
We can calculate real components of the curvature tensor 
(at symmetric points),    
which are directly related with low-energy scattering amplitudes, 
given by  
\beq
 && R_{A^i A^j A^k A^l}
    = f_{\pi}^2 {f_{RS}}^M f_{MUV}
     \delta^R_i \delta^S_j \delta^U_k \delta^V_l , \non
 && R_{B^i A^j B^k A^l}
    = - f_1 \vec{v}\,\dagg
    (Z_S \{Z_V,Z_U\} Z_R + Z_R \{Z_V,Z_U\}Z_S)\vec{v}
    \; \delta^R_i \delta^S_j \delta^U_k \delta^V_l \non
 && \hspace{2.3cm}
     - 4 g^2 (\delta_{ik} \delta_{jl} + \delta_{il} \delta_{kj}
            + \delta_{ij} \delta_{kl} ) , \non
 && R_{B^i A^j A^k A^l} =
    R_{A^i A^j B^k A^l} = 0 , \label{real-curvatures} 
\eeq
where we have defined $f_{\pi}^2 \defeq 2 f_1 v^2$.
We thus obtain low-energy (${\cal O}(p^2)$) 
scattering amplitudes of the NG and QNG bosons,  
by substituting Eqs.~(\ref{real-curvatures}) and (\ref{sym.of R}) 
to Eq.~(\ref{scat.amp.}).
We conclude that, at a symmetric point, 
there exist low-energy theorems of 
amplitudes which include only the NG bosons,
where higher derivatives of 
the arbitrary function cancel out,   
and they coincide with scattering amplitudes among NG bosons 
in non-supersymmetric theories 
on a symmetric space $G/H$ (see Eq.~(\ref{LET non-SUSY})).  
Amplitudes among only the QNG bosons 
coincide with those of the corresponding NG bosons 
by the K\"{a}hler conditions~(\ref{sym.of R}).  
Amplitudes for even number of the NG and QNG bosons 
depend on the second derivative of the arbitrary function. 

We would like to generalize these results to 
low-energy theorems at general points.
At non-symmetric points, 
some of the QNG bosons turn to NG bosons, 
corresponding to the fewer unbroken symmetry. 

\section{Non-symmetric points and supersymmetric vacuum alignment}
In the last section we have discussed 
low-energy theorems of NG and QNG bosons at 
a symmetric point. 
In supersymmetric low-energy theories, 
there can exist points with smaller unbroken symmertry 
in the same vacuum manifold, 
as a result of the supersymmetric vacuum alignment. 
In this section we discuss how this phenomenon occurs.  

\subsection{Non-symmetric points}
In maximal realizations, 
the number of the QNG bosons is equal to that of QNG bosons 
at symmetric points.
If we leave from the symmetric point by a $G$-action 
as $\vecv = g \vec{v} \;(g \in G)$, 
they are also symmetric points and the unbroken symmetry 
remains unchanged : $H\pri = g H g^{-1}\simeq H$.
The K\"{a}hler potential, the metric and the curvature tensor
do not change and the low-energy theorems do not change either.     
All of them are equivalent vacua.
The full target manifold is, however, constructed by 
$G^{\bf C}$-actions on $\vec{v}$.
If we move to another vacua by a $G^{\bf C}$-action, 
the unbroken symmetry $H$ varies depending on the choice of vacuum.
A $G^{\bf C}$-action on the symmetric point $\vec{v}$ is 
\beq
 \vecv = g_0 \vec{v} ,\hs{10}  
 g_0 \in G^{\bf C} \;.
\eeq
Complex unbroken subgroups at $\vec{v}$ and $\vecv$, defined by 
\beq
 \hat H \vec{v} = \vec{v} ,\hs{10} 
 \hat H\pri \vecv = \vecv  \;,
\eeq
are related by 
\beq
 \hat H\pri = g_0 \hat H {g_0}^{-1} \simeq \hat H : \hs{10}
 {K_M}\pri = g_0 K_M {g_0}^{-1} \;. \label{non-sym.unbr.}
\eeq
In this sense, the complex unbroken generators are equivalent 
at any point on the manifold .
Complex broken generators are also related by 
\beq
 {\cal G}^{\bf C}- \hat{\cal H}\pri 
 = g_0 ({\cal G}^{\bf C}- \hat{\cal H}) {g_0}^{-1} : \hs{10}
 {Z_R}\pri =  g_0 Z_R {g_0}^{-1} \;. \label{non-sym.br.}
\eeq
Since $G^{\bf C}$-orbits of $\vec{v}$ and $\vecv$ are 
homeomorphic to each other 
\beq
 G^{\bf C}/\hat H \simeq G^{\bf C}/\hat H\pri \;,
\eeq
the transformation of $G^{\bf C}$ 
is just an automorphism on the target manifold.
To be precise, (bosonic part of) the representatives of 
the complex cosets  
\beq
     \xi = \exp(i \ph^R Z_R) \in G^{\bf C}/\hat H  ,\hs{10}
 \xi\pri = \exp(i \ph^{\prime R} {Z_R}\pri) 
     \in G^{\bf C}/\hat H\pri
\eeq
are related by a right action of $G^{\bf C}$ through the relation
\beq
 \xi \vec{v} = \xi g_0^{-1} \cdot g_0 \vec{v} 
    = \left[\xi g_0^{-1}{\hat h}^{\prime -1}(\xi,g_0)\right] \vecv
    = \xi\pri \vecv  \;, \label{vxi} 
\eeq
as 
\beq
 \xi \pri = \xi g_0^{-1}{\hat h}^{\prime -1}(\xi,g_0) ,\hs{10}
 \hat h\pri \in \hat H\pri \;.  \label{r-action}
\eeq
This relation is sketched in Fig.~1.
\begin{figure}
 \epsfxsize=12cm
 \centerline{\epsfbox{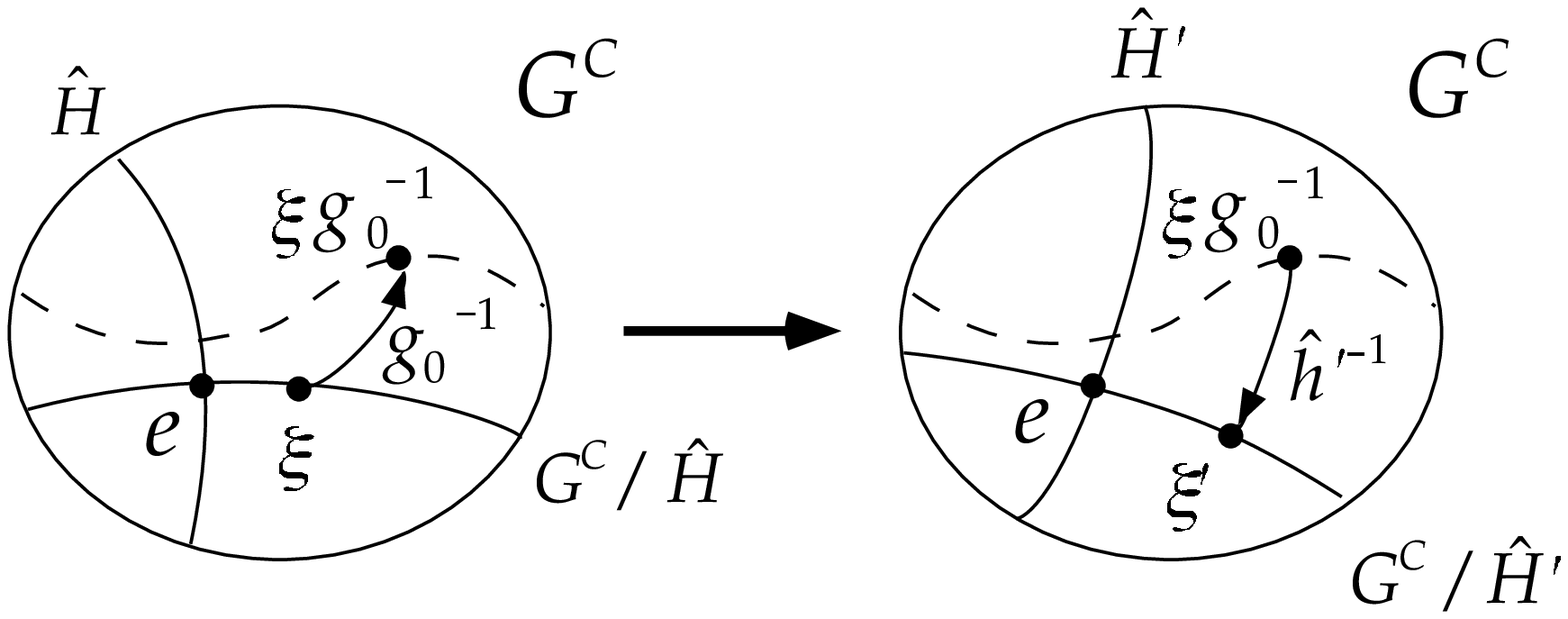}}
 \centerline{\bf Figure\,1}
\begin{footnotesize}
The complex unbroken symmetry $\hat H$ 
is transformed to $\hat H\pri$ in $G^{\bf C}$.
The coset representatives of $G^{\bf C}/\hat H$, $\xi$, 
which is expressed by a horizontal curve, 
are transformed to a broken curve by a right action of $g_0^{-1}$. 
To get transformed representatives of $G^{\bf C}/\hat H\pri$, 
$\xi\pri$, 
we need a local $\hat H\pri$ compensator from the right. 
\end{footnotesize}
\end{figure}

We can summarize these facts as follows: 
Suppose we choose a coordinate system 
whose origin is a symmetric point $\vec{v}$, 
and move to a non-symmetric point by an action of $g_0$  
and define a new coordinate system whose origin is $g_0 \vec{v}$. 
Then, unless $g_0$ belongs to the isometry $G$ of the metric, 
the origin of the new coordinate system $\ph\pri$ 
is no longer a symmetric point. 
The right action (\ref{r-action}) can be written explicitly as
\beq
 e^{i \ph\pri\cdot Z\pri} = e^{i \ph \cdot Z} {g_0}^{-1} 
    e^{-i u\pri(\xi,g_0)\cdot K\pri} ,\hs{10} 
 \hat h\pri = e^{i u\pri(\xi,g_0)\cdot K\pri} \in \hat H\pri \;, 
\eeq
where $u'$ is a function of $g_0$ and $\ph$. 
It can be rewritten as 
\beq
  e^{i \ph\pri\cdot Z} = {g_0}^{-1} e^{i \ph\cdot Z} 
  e^{-i u\pri(\xi,g_0)\cdot K} ,\hs{10} 
  e^{i u\pri(\xi,g_0)\cdot K} \in \hat H, 
\eeq
and if $g_0$ is restricted in $G$, 
it reduces to the ordinary left action (\ref{left-act.}). 
The right action does not change the K\"{a}hler 
potential from Eq.~(\ref{vxi}),
\beq
 \vec{v}\dagg \xi\dagg \xi \vec{v} 
  \to \vecvd \xi\prid \xi\pri \vecv = \vec{v}\dagg \xi\dagg \xi \vec{v}.
\eeq
We should again emphasize that  
this is just a coordinate transformation from 
the coordinate whose origin is a symmetric point $\vec{v}$
to one whose origin is a non-symmetric point $\vecv$,  
but not a symmetry.

\subsection{Supersymmetric vacuum alignment}
This subsection is devoted to an another interpretation, 
in terms of the group theory,  
about phenomena discussed in the last subsection, 
and then can be skipped. 
The compact subgroup $G$ is called a real form of $G^{\bf C}$.
The operation of $\;\cap {\cal G}$ 
on the complex algebra ${\cal G}^{\bf C}$ or its subalgebra
picks up Hermitian generators.
The real unbroken symmetry at the vacuum $\vec{v}$ 
is defined by 
\beq
 H = \hat H \cap G,  
\eeq
and at the vacuum $\vecv$ by 
\beq
 H\pri = \hat H\pri \cap G \neq H ,\hs{10} 
 (H') \subset (H) ,
\eeq
where $(\,\cdot\,)$ denotes an equivalence class by 
the $G$ action: 
$(H) = \{g H g^{-1}| g \in G\}$. 
Hence, at the non-symmetric point, 
the unbroken symmetry group becomes smaller than at 
the symmetric point~\cite{KS,KS2,LRM,Ni}. 
This phenomenon is called the  
``supersymmetric vacuum alignment''. 
It comes from the different embedding of 
$\hat H$ in $G^{\bf C}$ as in Fig.~2.\footnote{ 
BKMU called the embedding corresponding to $\vec{v}$ and $\vecv$ 
the ``natural embedding'' and 
the ``twisted embedding'', respectively~\cite{BKMU}. 
However they discussed the natural embedding only.   
Kotcheff and Shore called $\vec{v}$ and $\vecv$ 
the ``symmetric embedding'' and 
the ``non-symmetric embedding'', respectively, 
since they discussed the case when $G/H$ is a symmetric space and 
$G/H\pri$ is a non-symmetric space~\cite{KS}.
}
\begin{figure}
 \epsfxsize=5cm
 \centerline{\epsfbox{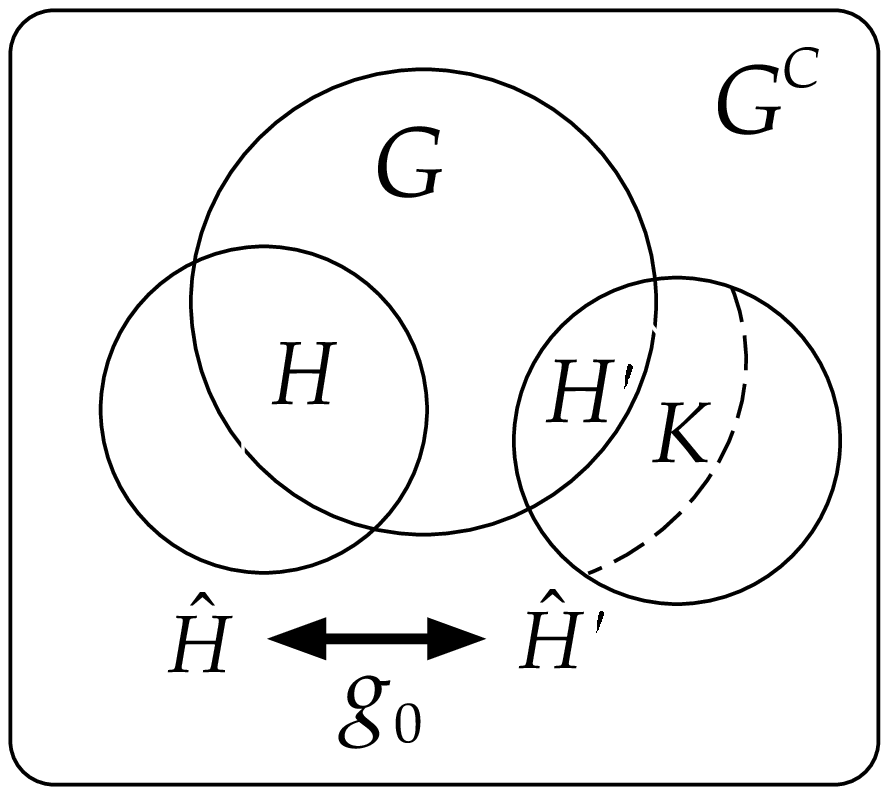}}
 \centerline{\bf Figure\,2}
\begin{footnotesize}
The large circle indicates the group $G$. 
The small circles denote 
the complex subgroups ${\hat H}$ and ${\hat H}\pri$.
${\hat H}\pri$ is the transform of ${\hat H}$ by $g_0$.
The real subgroups $H$ or $H\pri$ are defined as 
intersections of $G$ and $\hat H$ or ${\hat H}\pri$.
$K$ is the image of $H$ by the $g_0$ transformation.
In general $H\pri$ is a subset of $K$.
\end{footnotesize}
\end{figure}
In the case of $\hat H'$, 
$K \defeq g_0 H g_0 ^{-1} \;(\simeq H)$ 
is not a subgroup of $G$,
and the real unbroken symmetry is 
$K \cap G = H\pri \;(\neq H)$. 
At the non-symmetric point,
the generators ${\cal K} -{\cal H'}^{\bf C} \;
( \cap \;{\cal G} = \phi)$ 
are not Hermitian generators.
We call them ``pseudo-Borel generators''.\footnote{
A Borel algebra ${\cal B}$ ($\in \hat{\cal H}$) is defined as 
an algebra which satisfies, $[{\cal H},{\cal B}] \subset {\cal B}$.
Since ${\cal K} -{\cal H\pri}^{\bf C}$ does not satisfy 
this condition, it is not a Borel subalgebra. 
}
We show in the following sections that 
they correspond to NG bosons that appear at the non-symmetric point
where the unbroken symmetry becomes smaller.
The (real) $G$-orbits of $\vec{v}$ and $\vecv$ are
$G/H$ and $G/H\pri \;(\neq G/H)$, respectively. 
They are compact submanifolds of $G^{\bf C}/\hat H$ 
and parametrized by the NG bosons.
Other directions of the total target space $G^{\bf C}/\hat H$ are
non-compact and correspond to the QNG bosons.

\section{NG coset subspaces}
In this section, we discuss geometric structures of 
complex coset manifolds.  
To be specific we treat an simple example, the $O(N)$ model, 
but generalizations to other models are straightforward. 

\subsection{Typical example : $O(N)$ model}
In this subsection, we discuss the simplest example, 
the $O(N)$-model,
where the vacuum $\vec v$ is in the vector representation of 
$G=O(N)$: $\vec v \in V = {\bf R}^N$~\cite{Ni}.
We can complexify the $O(N)$ group by replacing 
the real vector with the complex vector: 
$V \to V^{\bf C} = {\bf C}^N$.
The generators of the $O(N)$ group are   
\beq
 &&{(T_{ij})^k}_l 
   = \1{i}({\delta_i}^k \delta_{jl} - {\delta_j}^k \delta_{il}) 
   = \pmatrix{  &  &  & \cr
              &  & i& \cr  
              &-i&  & \cr
              &  &  & \cr}  \; , 
\eeq
where only $(i,j)$ and $(j,i)$ elements are non-zero. 
They satisfy commutation relations and normalization conditions 
\beq
 &&[\; T_{ij} , T_{kl} \;] 
   = -i(\delta_{jk} T_{il} - \delta_{ik} T_{jl}
      - \delta_{jl} T_{ik} + \delta_{il} T_{jk}), \non
 && {\rm tr} (T_{ij} T_{kl}) = 
   2(\delta_{ik} \delta_{jl} - \delta_{il} \delta_{jk}) .
\eeq
For later convenience, we define
\beq
 X_i \defeq T_{Ni} \;,\; {X_i}\pri \defeq T_{N-1,i\pri}
 \;,\; X_{N-1} \defeq T_{N,N-1} \;\;(i,i\pri=1,\cdots,N-2)  \;.
\eeq

Let us classify the real and complex generators at 
1) a symmetric point and 2) a non-symmetric point.

\medskip
1) Symmetric points.\\ 
The vacuum vectors of symmetric points can be transformed 
by a $G$-action to
\beq
 \vec{v} = \pmatrix{0 \cr \vdots \cr 0 \cr v\cr} . \label{sym.point}
\eeq
We can immediately find a) the real unbroken algebra 
and b) the complex unbroken algebra.

1-a) Real unbroken Lie algebra.\\
Hermitian unbroken generators are $(N-1)\times (N-1)$ matrices 
which act the first $N-1$ components of 
(\ref{sym.point}), and others are broken generators:  
\beq
 \cal G 
 = \left(
   \begin{array}{ccc|c}
      &         &  &      \\
      &{\cal H} &  & * \\ 
      &         &  &      \\ \hline
      &  *      &  & 0  
   \end{array}
   \right) ,
\eeq
where ``$*$'' denote Hermitian broken generators. 
The Hermitian broken generators are of the form 
\beq
 X_i = T_{Ni} 
 = \left(
   \begin{array}{ccc|c}
              &        &        & \vdots \\
              & \ddots &        & i         \\ 
              &        &        & \vdots \\ \hline
       \cdots &  -i    & \cdots & 0  
   \end{array}
   \right) \in {\cal G} - {\cal H} \;\;\;(i = 1,\cdots ,N-1) ,
\eeq
where dots denote zero components and 
only $i$-th elements are nonzero. 
Thus the symmetry breaking pattern at symmetric points is 
$G = O(N) \to H=O(N-1)$.
A real target manifold, parametrized by 
NG bosons for this breaking, 
is a compact homogeneous manifold 
$O(N)/O(N-1)\simeq S^{N-1}$.

\bigskip
1-b) Complex unbroken Lie algebra.\\
To discuss the full target manifold, 
we must discuss by the comlexification of $G$. 
By the complexification, however, 
no new generator appears that leave the 
vacuum expectation value invariant at the symmetric point. 
Broken and unbroken generators are simply 
\beq
\cases{
 Z_R = X_i \in {\cal G}^{\bf C} - \hat{\cal H}, \cr
 K_M = H_a \in \hat{\cal H} .}  
\eeq
All $Z_R$ are Hermitian generators. 
Chiral superfields, whose bosonic parts are 
coset coordinates, proportional 
to Hermitian generators are called the ``mixed-type superfield''. 
A real part of a mixed type superfield is a NG boson 
while an imaginary part corresponds to a {\it QNG boson}.
Since numbers of the NG and QNG bosons are 
both $N-1$ in this case, 
this nonlinear realization is called the maximal realization.

\medskip
2) Non-symmetric points.\\
We discuss the symmetry breaking in general points. 
We move the vacuum expectation value $\vec v$ to $\vecv$ 
by the following element of $G^{\bf C}$:   
\beq
 g_0 &=& \exp(i \th X_{N-1}) \non 
 &=& \left(
   \begin{array}{c|cc}
               \LARGE{1}& \LARGE{0}&          \\ \hline
                        & \cos \th & -\sin\th \\ 
               \LARGE{0}& \sin \th &  \cos\th  
   \end{array}
   \right)
 = \left(
   \begin{array}{c|cc}
               \LARGE{1}& \LARGE{0}   &             \\ \hline
                        & \cosh \tht  & -i\sinh\tht \\ 
               \LARGE{0}& i\sinh \tht &  \cosh\tht  
   \end{array}
   \right) \;\;\in G^{\bf C}  \;,
\eeq
where $\th \defeq i \tht$ is a pure imaginary angle.
Although we have chosen the rotation by $X_{N-1}$, 
rotations by other broken generators are equivalent,  
since they can be transformed by an action of $G$ to each other.
Then the vacuum vector at the non-symmetric points can be written, 
without loss of generality, as
\beq
 \vecv 
 = g_0 \vec{v} 
 = \pmatrix{0 \cr \vdots \cr 0 \cr -i v \sinh \tht \cr  v \cosh \tht \cr} 
 = \pmatrix{0 \cr \vdots \cr 0 \cr\alpha \cr \beta\cr} , \label{vpri}
\eeq
where we have defined
\beq
 \cases 
 {\beta \defeq v \cosh \tht \;:\; {\rm real} \cr
  \alpha \defeq - i v \sinh \tht \;:\; {\rm pure \;imaginary}, 
  } \label{noncom.}
\eeq
and these satisfy ${\beta}^2 + \alpha^2 = v^2$.
By these constants, $g_0$ and its inverse can be written as
\beq
 g_0 
= \left(
   \begin{array}{c|cc}
               \LARGE{1}& \LARGE{0}   &            \\ \hline
                        & \beta / v   & \alpha / v \\ 
               \LARGE{0}& -\alpha / v & \beta / v  
   \end{array}
   \right) ,\hs{10} 
 {g_0}^{-1} 
= \left(
   \begin{array}{c|cc}
               \LARGE{1}& \LARGE{0}   &            \\ \hline
                        & \beta / v   & -\alpha / v \\ 
               \LARGE{0}& \alpha / v  & \beta / v  
   \end{array}
   \right) .
\eeq
The magnitudes of vacuum expectation values are
\beq
 \vecvd \vecv = \beta^2 - \alpha^2 
 = \beta^2 + \alpht^2 \defeq {v\pri}^2 \;,\\
  \vec{v}^{\prime 2} = \vec{v}^2 
  = \beta^2 + \alpha^2 = \beta^2 - \alpht^2 = v^2 \;.
\eeq
We can find from Eq.~(\ref{vpri}) that 
the non-compact directions are hyperbolic as in Fig.~3.
The full target space is a spheroidal hyperboloid 
and the compact coset $G/H$ is embedded in the symmetric point. 
(The compact coset $G/H\pri$, at non-symmetric point, 
is discussed in the next subsection.)
\begin{figure}
 \epsfxsize=7.5cm
 \centerline{\epsfbox{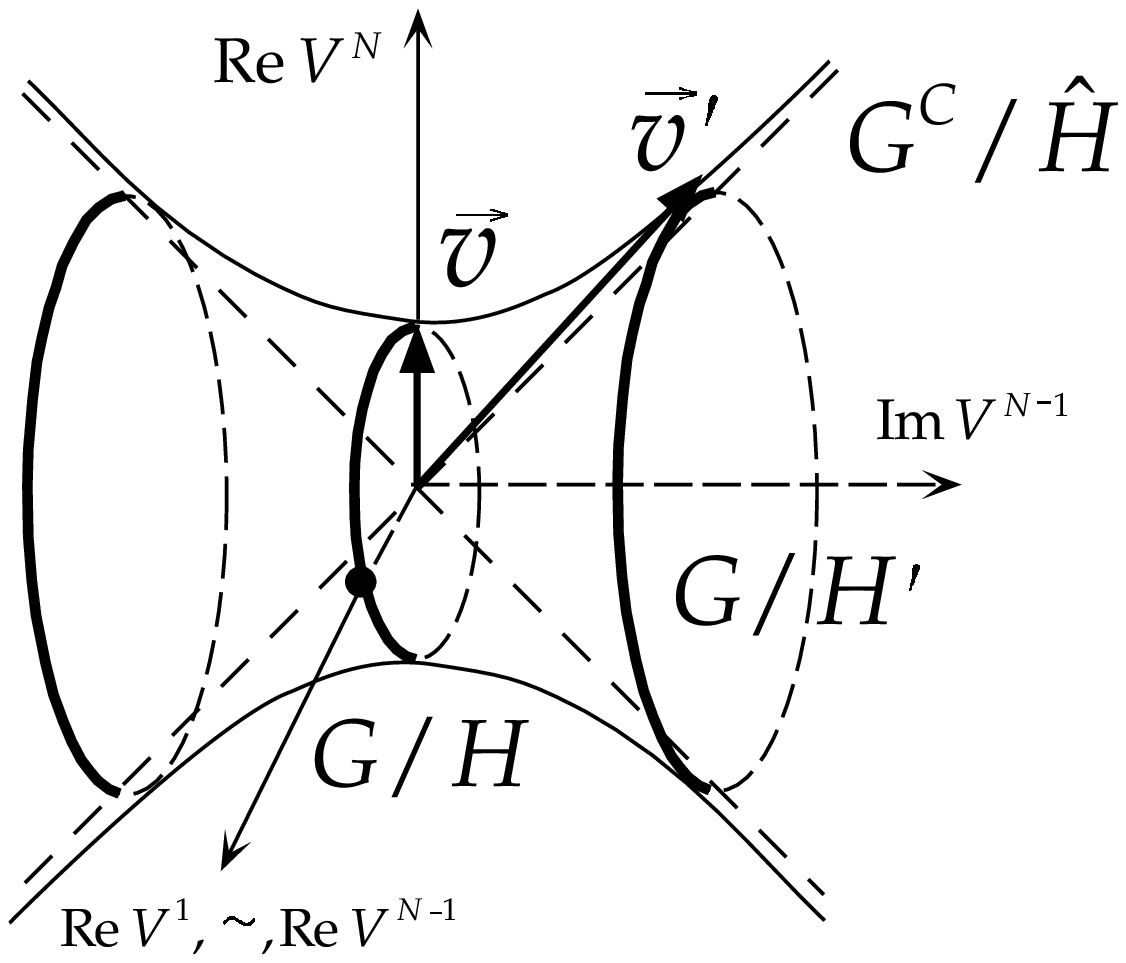}}
 \centerline{\bf Figure\,3}
\begin{footnotesize}
The vertical axis is a real part of $V^N$, 
other real parts are written as the axis to this side 
and the right axis is an imaginary part of $V^{N-1}$, 
parametrized by $\alpht$. 
The NG coset manifold at the symmetric point, $G/H$, 
is written as a vertical circle of a radius $v$, at the center.  
The NG coset manifold at the non-symmetric point, 
$G/H\pri$, is written as two vertical circles of a radius $\beta$, 
and both circles are connected, by a $G$ action, through 
other imaginary directions of $V$. (See also Fig.~4, below.)  
\end{footnotesize}
\end{figure}

\bigskip
Let us discuss real and complex Lie algebras at 
non-symmetric points.\\ 
2-a) Real Lie algebra.\\
At non-symmetric points, 
the whole generators can be divided into 
real unbroken algebra ${\cal H}\pri$ and 
real broken generators as  
\beq
 \cal G 
 = \left(
   \begin{array}{ccc|cc}
       &                  & &   &   \\ 
       & \LARGE{{\cal H}'}& & * & * \\ 
       &                  & &   &   \\ \hline
       & *                & & 0 & * \\ 
       & *                & & * & 0  
   \end{array}
   \right)  \;,
\eeq
where ``$*$'' denote broken generators. 
The broken generators can be written explicitly as
\beq
&& {X_i}'
 = \left(
   \begin{array}{ccc|cc}
               &       &       &\vdots & \vdots  \\
               &\ddots &       & i     & 0 \\ 
               &       &       &\vdots & \vdots \\ \hline
        \cdots & -i    &\cdots &0 & 0 \\ 
        \cdots &  0    &\cdots &0 & 0  
   \end{array}
   \right),\hs{10}
 X_i
 = \left(
   \begin{array}{ccc|cc}
              &        &       &\vdots &\vdots   \\
              & \ddots &       & 0     & i \\ 
              &        &       &\vdots &\vdots \\ \hline
       \cdots &  0     &\cdots &0 & 0 \\ 
       \cdots &  -i    &\cdots &0 & 0  
   \end{array}
   \right) ,\non
&& X_{N-1}
 = \left(
   \begin{array}{ccc|cc}
      &       & &       &   \\
      &\ddots & &\vdots &\vdots   \\ 
      &       & &       &   \\ \hline
      &\cdots & & 0     & i \\ 
      &\cdots & & -i    & 0  
   \end{array}
   \right) 
  \in {\cal G}-{\cal H}\pri \hs{10} (i=1,\cdots,N-2) \;,
\eeq
where dots denote zero components, 
and non-zero elements in the first two equations 
are $i$-th components. 
The symmetry breaking pattern at non-symmetric points turns out  
$G = O(N) \to H\pri = O(N-2)$, 
which is smaller than symmetric points.
A real target manifold, parametrized by NG bosons,  
is a compact homogeneous manifold 
$G/H\pri = O(N)/O(N-2)$, 
which is larger than one of the symmetric point, $G/H$. 
(See Fig.~3.)
Namely we have more NG bosons at non-symmetric points   
than at symmetric points.
These newly emerged NG bosons must come from the QNG bosons,
since the dimension of the full target manifold has to be unchanged.
There is only one QNG boson because the number of 
the NG bosons is $2N-3$ (and the total number is $2N-2$). 
In the next subsection, 
we show how these different compact coset manifolds are 
embedded in the full manifold and 
how some of the QNG bosons change to the NG bosons at 
non-symmetric points.
Before doing it, we investigate the complex symmetry 
at non-symmetric points, 
which give us the key point to understand such phenomena.

\medskip
2-b) Complex Lie algebra.\\
Complex broken and unbroken generators 
at non-symmetric points $\vecv$  
can be immediately calculated by using  
Eqs.~(\ref{non-sym.br.}) and (\ref{non-sym.unbr.}), to yield 
\beq
 {Z_R}\pri 
 &=& g_0 Z_R g_0^{-1} \non
 &=& \cases {g_0 X_i g_0^{-1}
             = {\alpha \over v} {X_i}\pri 
             + {\beta \over v} X_i \defeq {Z_I}\pri \cr
             g_0 X_{N-1} g_0^{-1} = X_{N-1} \defeq Z_{N-1}^{\;\prime}
             } 
  \;\;\in {\cal G}^{\bf C}- \hat{\cal H}\pri \;,
 \\
 {K_M}\pri 
 &=& g_0 K_M g_0^{-1} \non
 &=& \cases {g_0 X_i\pri g_0^{-1}
             = {\beta \over v} {X_i}\pri 
             - {\alpha \over v} X_i \defeq {B_I}\pri \cr
             g_0 {H_a}\pri g_0^{-1} = {H_a}\pri \in {\cal H}\pri
             } 
  \;\; \in \hat{\cal H}\pri  \; ,
\eeq
where ${Z_I}\pri$ and ${B_I}\pri$ $(I\pri=1,\cdots,N-2)$ 
can be explicitly written as
\beq
 &&{Z_I}'
 = \left(
   \begin{array}{ccc|cc}
            &           &       &\vdots     &\vdots   \\
            &\ddots     &       &i \alpha/v & i\beta/v \\ 
            &           &       &\vdots     &\vdots   \\ \hline
     \cdots &-i\alpha /v&\cdots & 0         & 0       \\ 
     \cdots &-i\beta /v &\cdots & 0         & 0  
   \end{array}
   \right)  ,\non
 &&{B_I}'
 = \left(
   \begin{array}{ccc|cc}
             &           &       &\vdots    &\vdots      \\
             &\ddots     &       &i \beta/v & -i\alpha/v \\ 
             &           &       &\vdots    & \vdots     \\ \hline
      \cdots &-i\beta /v &\cdots & 0        & 0          \\ 
      \cdots &i\alpha /v &\cdots & 0        & 0  
   \end{array}
   \right) \; . 
\eeq
We can classify these broken generators 
to pure-types or mixed-types as follows.
First of all, 
the broken generator ${Z_{N-1}}'$ corresponds to 
a mixed-type superfield,  
since ${Z_{N-1}}'$ is a Hermitian generator. 
A real and a imaginary parts of a scalar component of 
a chiral superfield generated by ${Z_{N-1}}'$ 
is a NG boson and a QNG boson, respectively.  
On the other hand, 
all other generators ${Z_I}'$ generate 
pure-type chiral superfields,
where both scalar components are NG bosons, 
since they are non-Hermitian generators. 

We can count numbers of the NG and QNG bosons as 
$2N-3$ and $1$, respectively, 
without using the fact that 
the total number of the NG and QNG bosons does not change.

\subsection{Embedding of NG cosets $G/H$ and $G/H\pri$}
In this subsection,
we show how the different cosets 1) $G/H$ and 2) $G/H\pri$ 
are embedded into symmetric and non-symmetric points by 
using the Shore's procedure~\cite{Sh,KS}.
We can obtain coset representatives of NG bosons 
$G/H$ ($G/H\pri$) by putting all QNG bosons zero 
in the complex representative $\xi$ 
of the full complex coset $G^{\bf C}/\hat H$, 
at symmetric (non-symmetric) points.
In the cases when there are pure-type superfields,  
we need a local $\hat H$-transformation from the right. 

1) Embedding of $G/H$ at symmetric points.\\
Since we do not need a local $\hat H$-transformation, 
we can obtain the representative of $G/H$ 
by simply putting all QNG bosons zero~\cite{KS}:  
\beq
  \xi|_{{\rm QNG}=B^i=0} 
  = e^{i\phi \cdot X} \in G/H   \;,
\eeq
where fields $\phi = \{A^i\} \;\;(i=1,\cdots,N-1)$ 
are NG bosons at symmetric points. 

\medskip
2) Embedding of $G/H'$ at non-symmetric points.\\
The representative of the complex coset 
at non-symmetric points is
\beq
 \xi(\ph) 
 = e^{i\ph \cdot Z\pri}  
 = \exp i \left[ \ph^i 
  \left({\beta\over v} X_i + {\alpha\over v} {X_i}'\right)
            + \ph^{N-1} X_{N-1}\right]  
  \in G^{\bf C}/\hat H\pri \;, 
\eeq
where we have used a character $\ph$ as a coordinate.  
Since there are pure-type broken generators ${Z_I}'$, 
we need a local $\hat H$-transformation 
\beq
 \zeta\pri (\ph,\ph^*) 
 = \exp (i d (\ph,\ph^*)\cdot {B}') \in \hat H\pri
\eeq
from the right:  
\beq
 \xi (\ph) \to &&  \non
 \hat{\xi}(\hat A,\hat B) 
 &=& \xi(\ph) \zeta\pri (\ph,\ph^*)\non
 &=& \exp i\left[ \hat\ph^i \left({\beta\over v} X_i 
                       + {\alpha\over v} {X_i}'\right)
  + \hat d^i(\hat\ph,\hat\ph^*)
       \left( -{\alpha \over v}X_i + {\beta \over v} {X_i}'\right)
  + \hat\ph^{N-1} X_{N-1} \right] \non
 &=& \exp i\left[a^i X_i + b^i {X_i}\pri \
            + (\hat A^{N-1} + i \hat B^{N-1})X_{N-1}\right] \;,
  \label{emb.G/Hp}
\eeq
where $\hat \ph^i = \hat A^i + i \hat B^i$ are transformed fields   
whose relation to $\ph$ is obtained below, Eq.~(\ref{rel.ph}),   
$\hat d^i$ is a function of $\hat \ph$ and $\hat \ph^*$ 
whose relation to $d^i$ is also obtained below, Eq.~(\ref{d-hat-d}), 
and $a^i$ and $b^i$ are scalar fields. 
We can chose the function $\hat d$ (or $d$) such that 
scalar fields $a$ and $b$ become real: 
\beq
 \hat d^i (\hat\ph,\hat\ph^*) 
  = - {i \over 2 \tilde \alpha \beta} 
      (v^{\prime 2} \hat \ph^i - v^2 \hat \ph^{*i}), \hs{10}
 a^i 
 = \hat A^i \;{v \over \beta},\hs{10}
 b^i 
 = \hat B^i \;{v \over \tilde \alpha}.   
 \label{dab}
\eeq 
Since real scalar fields $\hat A^i$ and $\hat B^i$ 
are proportional to Hermitian generators 
$X_i$ and $X_i'$, respectively in the exponential,   
they parameterize compact directions of the target manifold.  
This is why both $\hat A^i$ and $\hat B^i$ are NG bosons, 
and $\hat\Ph^i$ can be considered pure-type superfields. 
On the other hand, since $\hat A^{N-1}$ and $\hat B^{N-1}$ are 
proportional to Hermitian and anti-Hermitian generators 
on the exponential,  
they parameterize compact and non-compact directions of 
the target space, respectively. 
Hence $\hat A^{N-1}$ and $\hat B^{N-1}$ are NG and QNG bosons, 
and then $\hat\Ph^{N-1}$ is a mixed-type superfield.
Then the superfields $\Ph^i$ and $\Ph^{N-1}$, 
before the $\hat H$-transformation,  
turn out to be pure-type and mixed-type superfields, respectively, 
since they coincide with $\hat \Ph^i$ and $\hat\Ph^{N-1}$ 
at the linear order, as shown below, Eq.~(\ref{rel.ph}). 
 
We can obtain the real representative of the coset $G/H\pri$ 
by putting all QNG bosons zero: 
\beq
 && \hat{\xi}|_{{\rm QNG} = \hat B^{N-1} = 0} 
   = e^{i\phi' \cdot X} \in G/H\pri \;,
\eeq
where fields $\phi' = \{ a^i , b^i , \hat A^{N-1}\}$ 
are NG bosons at non-symmetric points. 
To understand why the compact coset manifold $G/H'$ 
at non-symmetric points is 
larger than $G/H$ at symmetric points, see Figs.~4 and 5. 
At symmetric points, 
there are $N-1$ non-compact directions, 
while they change to one non-compact direction 
and $N-2$ compact directions 
parametrized by newly emerged NG bosons.\footnote{
We give a comment on the NG submanifold 
at non-symmetric points, $G/H\pri$.
It can be considered as 
a $H/H\pri \simeq S^{N-2}$ fiber bundle 
over a base manifold, $G/H \simeq S^{N-1}$. 
By bringing $\vecv$ to $\vec{v}$, the fiber shrinks but 
the base remains at finite size (radius $v$). 
}
\begin{figure}
  \epsfxsize=7.5cm
  \centerline{\epsfbox{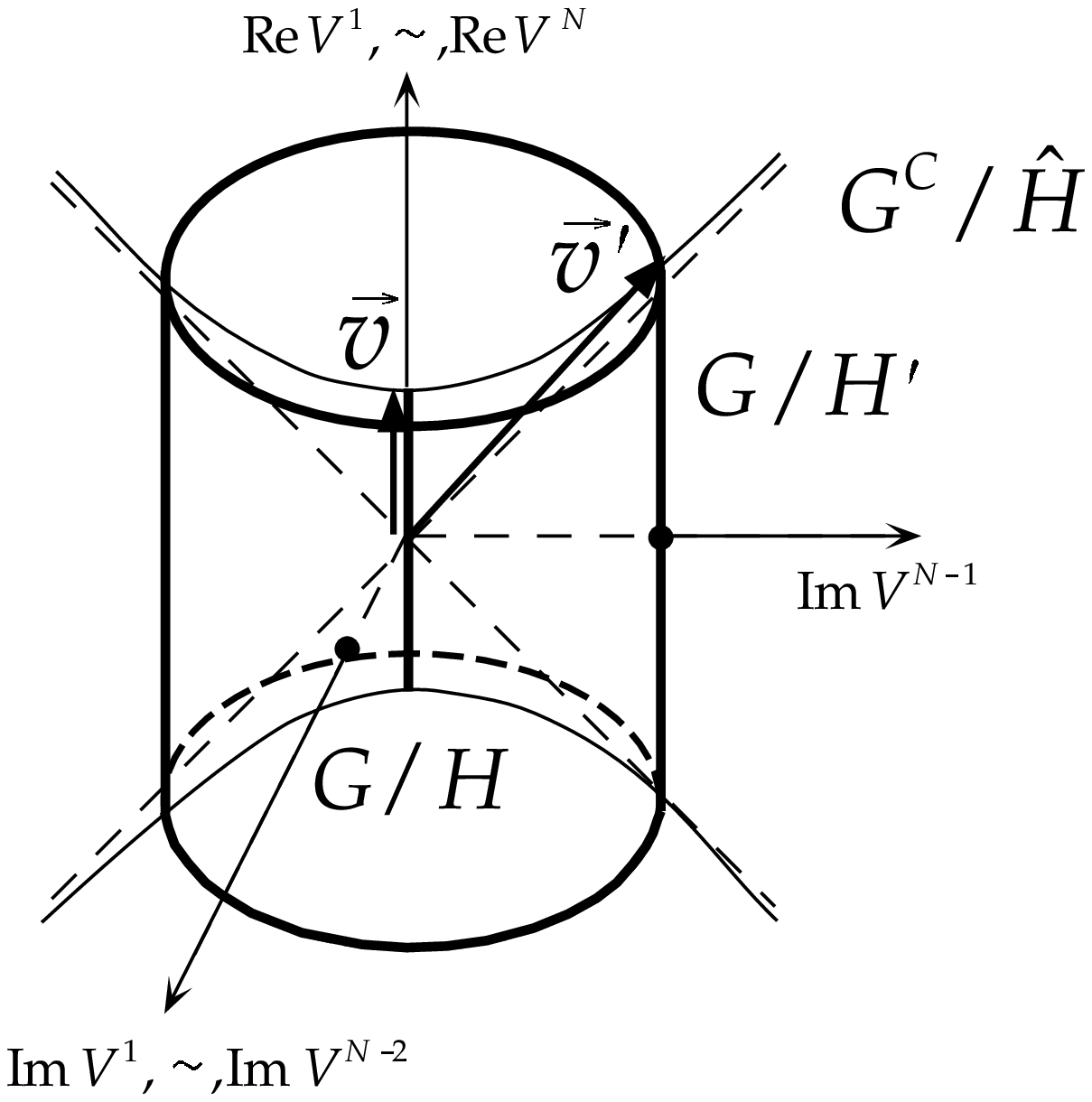}}
  \centerline{\bf Figure\,4}
\begin{footnotesize}
In Fig.~4, the imaginary directions of $V$ are written as 
a horizontal plane, and real directions are 
written as a vertical line.
$G/H$ is written as a segment at the center 
and $G/H\pri$ is written as a cylinder enclosing the $G/H$. 
Newly emerged NG bosons are written 
as a horizontal circle of a radius $\alpht$, 
which is just a $H$-orbit of the non-symmetric vacuum $\vecv$.
\end{footnotesize}
\bigskip

  \epsfxsize=5cm
  \centerline{\epsfbox{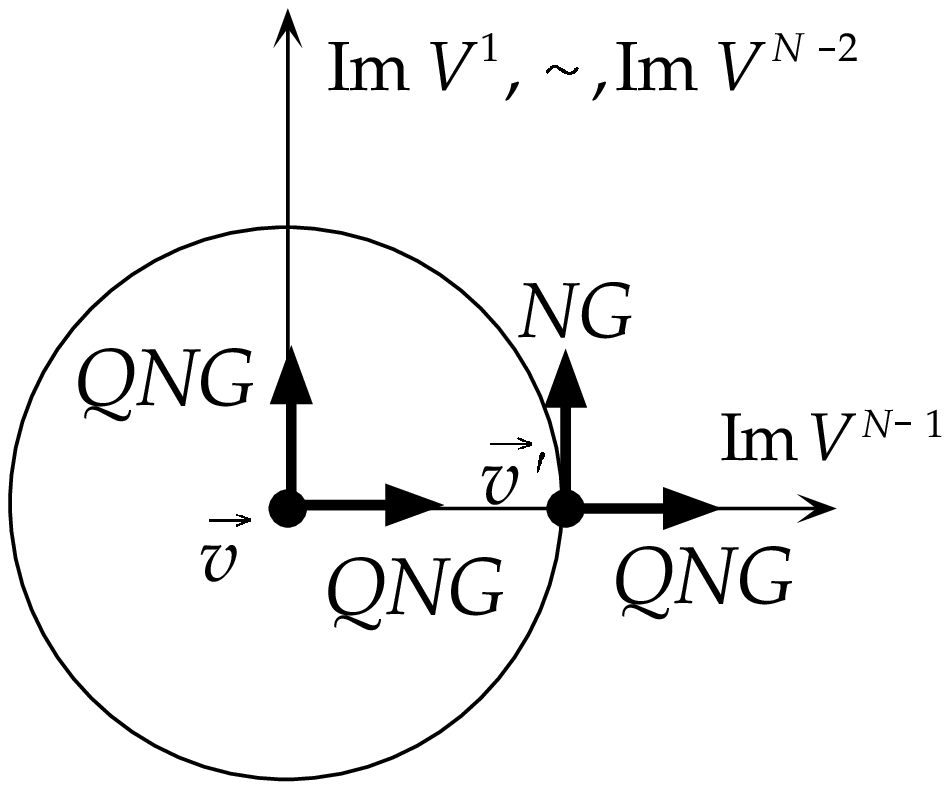}}
  \centerline{\bf Figure\,5}
\begin{footnotesize}
View from the top of Fig.~4. 
It can be seen that most of QNG bosons at the symmetric point 
change to newly emerged NG bosons at the non-symmetric point, 
with the total number of massless bosons been unchanged. 
\end{footnotesize}
\end{figure}

Next we obtain the relation of the fields $\ph$ and $\hat \ph$ 
(or $\Ph$ and $\hat \Ph$). 
In general, the first equation of Eq.~(\ref{emb.G/Hp}) 
can be written explicitly as 
\beq
  e^{i \ph \cdot Z} e^{i d(\ph,\ph^*) \cdot K} 
= e^{i(\hat \ph \cdot Z + \hat d (\hat\ph,\hat\ph^*) \cdot K)}  
 ,\hs{10}  
 \zeta \pri(\ph,\ph^*) 
   = e^{i d(\ph,\ph^*)\cdot K} \in \hat H\pri 
 \;, \label{local_zeta}
\eeq
where $d$ is a function of $\ph$ and $\ph^*$. 
By using the Baker-Campbell-Hausdorff formula on the left-side,  
we obtain relations 
\beq
 && \hat{\ph}^R = \ph^R 
   + \1{2} {f_{MS}}^R \ph^S d^M + \cdots ,\\
 && \hat d^M = d^M 
  + \1{12} {f_{NR}}^T {f_{TS}}^M \ph^R \ph^S d^N + \cdots .
  \label{d-hat-d}
\eeq
Note that we have used only the fact that 
$G^{\bf C}/\hat H$ is a symmetric space, 
and the result is model-independent. 
By these equations, 
two coordinates $\hat{\ph}$ and $\ph$ are related by 
\beq
 \hat{\ph} = \ph + O(\ph^2,\ph\ph^*), \label{rel.ph}
\eeq
and coincide to each other at the first order.\footnote{ 
Note that this transformation is not holomorphic and 
so the complex structures of the two coordinates 
$\ph$ and $\hat \ph$ are different.
For our purpose to calculate low-energy theorems, 
the difference can be neglected because the curvature tensors in
both coordinates coincide at $\ph =0$.
} 
This is why $\Ph$ and $\hat \Ph$ coincide to each other 
at the linear level and their identifications to  
pure- or mixed-type superfields coincide.

\section{Low-energy theorems at general points}
In this section, we discuss low-energy theorems 
at non-symmetric points.
We elaborate on the $O(N)$ model as an example, 
but generalizations to other models are straightforward.
  
\subsection{Some formula for the $O(N)$ model}
Before discussing low-energy theorems, 
we give comments for a ``linear'' description of the model. 
The invariant Lagrangian can be written as 
\beq
 {\cal L} 
 = \int d^4\th \vec{\phi}\dagg \vec{\phi} 
  + \left(\int d^2\th W(\phi)+ (\rm conj.) \right) , 
  \label{lin.lag.}
\eeq
where $\vec{\phi}(x,\th,\thb)$ consists of 
chiral superfields belonging to a linear representation of $G$. 
$W(\phi)$ is a $G$-invariant superpotential,  
and it is actually $G^{\bf C}$-invariant due to its holomorphy.
For the $O(N)$ model, its candidate is 
\beq
 W(\phi) = g \phi_0 ( \vec{\phi}^2 - a^2 ) \;, \label{superpot.}
\eeq
where $\phi_0$ is an additional $G$-singlet field 
and $g$ is a coupling constant. 
$\phi_0$ tends to a non-dynamical auxiliary field in 
the heavy mass limit ($g \to \infty$). 
In this limit we can eliminate $\phi_0$ by its equation of motion, 
which gives an F-term constraint among dynamical fields $\phi^i$. 
For the $O(N)$ model, 
the constraint is $\vec{\phi}^{\,2} - a^2 = 0$. 
The K\"{a}hler potential may suffer from 
a quantum correction, 
with preserving the global symmetry $G$. 
We thus obtain a nonlinear K\"{a}hler potential 
for NG chiral superfields, 
\beq
 K(\Ph,\Phd) 
 = f(\vec{\phi}^{\,\dagger}\vec{\phi})|_{\rm F} 
 = f(\vec{v}\dagg \xi\dagg \xi \vec{v} ) \;,\;
\eeq
where F denotes an F-term constraint. 
We have used a relation between linear superfields and NG superfields, 
$\vec{\phi}|_{\rm F} = \xi \vec{v}$.
This recovers Eq.~(\ref{A type}).  

If we restrict the problem to the $O(N)$ model, 
geometric quantities can be calculated by 
solving the constraint $\vec{\phi}^2 = a^2$ explicitly 
as $\phi^N = \sqrt{a^2 - \sum_{i=1}^{N-1}(\phi^i)^2}$. 
However we discuss in the coset formalism 
which can be generalized to other models straightforwardly.

\medskip
To obtain geometric quantities in the coset formalism,  
we need expectation values of broken generators, 
sandwiched by the vacuum vector $\vecv$. 
Let us calculate them first. 
We use indices 
$R,S,\cdots = 1, \cdots, N-1$ and $I,J,\cdots = 1, \cdots, N-2$.
We omit primes except for $\vec{v}'$. 
By noting that ${Z_{N-1}}\dagg = Z_{N-1}$ and ${Z_I}\dagg \neq Z_I$, 
we calculate products of one complex generator on 
vacuum expectation values, given by   
\beq
 Z_I \; \vecv 
  =  \pmatrix{0 \cr \vdots \cr i v \cr \vdots \cr 0},\hs{10} 
 Z_{N-1} \; \vecv 
  =  \pmatrix{0 \cr \vdots \cr \vdots 
              \cr i\beta \cr -i\alpha},\hs{10}
 {Z_I}\dagg \; \vecv =  
  {{v\pri}^2 \over v^2} 
   \pmatrix{0 \cr \vdots \cr i v \cr \vdots \cr 0}, 
\eeq
where the $I$-th elements are nonzero in 
the first and the third equations. 
Products of two complex generators on 
vacuum expectation values are also given by 
\beq
 && Z_R Z_S \; \vecv  = \delta_{RS} \vecv, \hs{10}
   {Z_I}\dagg Z_J \; \vecv = \delta_{I^*J} \vecvs  \;,\non
 &&{Z_I}\dagg Z_{N-1} \; \vecv  
  = {c^2 \over {v\pri}^2} {Z_I}\dagg \;\vecv ,\hs{10}
  Z_{N-1} {Z_I}\dagg \; \vecv = 0  \;,
\eeq
where we have defined 
$c^2 \defeq 2 \sqrt{v^{\prime\,4} - v^4} = 2 \tilde{\alpha}\beta$. 
We define convenient notations
\beq
 \big< R \cdots \big> \defeq \vecvd Z_R \cdots \vecv ,
 \hs{10} R\dagg \defeq {Z_R}\dagg .  \label{notations}
\eeq
Then expectation values of one to four generators 
can be calculated, to yield 
\beq
 && \big<I \big> = 0 ,\hs{36}
    \big<N-1 \big> = c^2 \;,\non
 && \big<RS\big> = {v\pri}^2 \delta_{RS} ,\hs{23.5}
    \big<I\dagg J\big> = v^2 \delta_{I^*J} \;,\non
 && \big<IJK\big> = 0, \hs{30}
    \big<I\dagg JK\big> = 0 \;,\non
 && \big<N-1,IJ\big> = c^2 \delta_{IJ} \;,\hs{13}
    \big<IJ,N-1\big> = 0 \;,\non
 && \big<*I,N-1\big> = 0 \;,\hs{19}
    \big<*,N-1,I\big> = 0 \;,\non
 && \big<I,N-1,N-1\big> = 0 \;,\hs{10}
    \big<I\dagg,N-1,N-1\big> = 0 \;,\non
 && \big<N-1,I,N-1\big> = 0 \;,\hs{10}
    \big<N-1,N-1,N-1\big> = c^2  \;, \non
 && \big<R\dagg S\dagg UV\big>  
 = {v'}^2 \delta_{R^*S^*} \delta_{UV}  \;, \label{vevs}
\eeq
where ``$*$'' denotes an arbitrary generator.
These quantities are needed for the calculation of 
the curvature tensor. 
They can be generalized to other models 
straightforwardly.

\subsection{Geometric quantities and low-energy theorems}
We can calculate geometric quantities of 
the $O(N)$ model by using the formulas obtained 
in the last subsection.
In this section, 
we consider the most simple K\"{a}hler potential, 
$K=f(x)=x$. 
The general case is discussed in Appendix B. 

First of all the metric is given by (we omit prime on $Z_R$)
\beq
 &&g_{ij^*} = \del_i \del_{j^*} K
            = G_{RS^*} E^R_i (E^S_j)^* ,\hs{10} 
 G_{RS^*} 
 = \vecvd {Z_S}\dagg \xi\dagg \xi Z_R \vecv,
\eeq
where $G_{RS^*}$ is called an auxiliary metric. 
The auxiliary metric at the point $\ph=0$ becomes
\beq
 G_{RS^*}|_{\ph=0} = \big<S\dagg R\big> 
 = \pmatrix { v^2 \delta_{IJ^*} & 0 \cr
            0 & {v\pri}^2 }   \;.\label{aux.metric}
\eeq
A vielbein and a ${\hat H}$-connection 
at the point $\ph=0$ are given by 
\beq
 E^R_i|_{\ph=0} = \delta^R_i \;,\hs{10}  
 W^M_i|_{\ph=0} = 0,
\eeq
respectively, and the differentiation of the vielbein 
with respect to the coordinate   
can be calculated, to yield 
\beq
 \del_j E^R_i|_{\ph=0} = 0  \;.
\eeq 
Let us calculate a complex curvature   
\beq
 R_{ij^*kl^*} = \del_i\del_{j^*}\del_k \del_{l^*}K
 - g^{mn^*}(\del_{j^*}\del_m \del_{l^*} K) 
           (\del_i \del_k \del_{n^*} K)  \;,\label{curvature}
\eeq
which is crucial to low-energy theorems.
The complex curvature on the origin $\ph=0$ of $G^{\bf C}/\hat H$ 
can be calculated by Eq.~(\ref{vevs}), to yield
\beq
 R_{ij^*kl^*}|_{\ph=0}  
 &=& [ \big<S\dagg V\dagg UR\big> 
  - G^{XY^*}|_{\ph=0} \big<X\dagg SV\big>^* \big<Y\dagg RU\big> ] 
   \delta^R_i (\delta^S_j)^* \delta^U_k (\delta^V_l)^*  \non
 &=&  {v^4 \over {v\pri}^2} \delta_{ik}\delta_{j^*l^*} \;. 
  \label{comp.curvature}
\eeq
The first line is for 
general symmetric manifolds $G^{\bf C}/\hat H$ with one vacuum expectation value, 
and the second line is for the $O(N)$ model.
If we rescale fields so that 
the metric (\ref{aux.metric}) becomes the Knonecker's delta, 
components of the curvature tensor become
\beq
 R_{ij^*kl^*}|_{\ph=0} 
 = \cases{
    1/{v\pri}^2  \delta_{ik}\delta_{j^*l^*} \; 
      \hspace{0.2cm}\mbox{ when } i,j,k,l = 1,\cdots,N-2, \cr
    v^2/{v\pri}^4 \delta_{ik}\delta_{j^*l^*} \; 
      \mbox{ when only two indices are $N-1$-th}, \cr
    v^4/{v\pri}^6 \delta_{ik}\delta_{j^*l^*} \;
    \mbox{ when } i,j,k,l = N-1 .
   } 
\eeq 
\ From these equations, 
we can calculate real components of the curvature 
in the rescaled coordinate given by 
$(i,j,k,l=1,\cdots,N-2)$ 
\beq
&& R_{A^iA^jA^kA^l} 
  = 2{ 1 \over {v\pri}^2}
    (\delta_{ik}\delta_{jl}-\delta_{il}\delta_{jk}),\non
&& R_{B^iA^jB^kA^l} 
  = - 2 {1 \over {v\pri}^2}
        (\delta_{ik}\delta_{jl} + \delta_{il}\delta_{jk}) ,\non
&& R_{B^iA^jA^kA^l} = R_{A^iA^jB^kA^l} = 0 ,\non
&& R_{A^{N-1}A^jA^{N-1}A^l} = 2{v^2\over {v\pri}^4}\delta_{jl},\non
&& R_{A^{N-1}A^{N-1}A^kA^l} = 0 ,\non
&& R_{B^{N-1}A^jB^{N-1}A^l} =-2{v^2\over {v\pri}^4}\delta_{jl},\non
&& R_{B^{N-1}A^{N-1}B^kA^l} = 0 ,\non
&& R_{B^{N-1}A^{N-1}B^{N-1}A^{N-1}} = -4 {v^4 \over {v\pri}^6} \;,  
 \label{comp.curvature2}
\eeq
where all quantities are evaluated at $\ph=0$. 

Let us discuss low-energy theorems.
As discussed in Appendix A, 
components of the curvature tensor in an arbitrary coordinate  
and in normal coordinates coincide to each other 
in this order. 
Hence, by substituting these equations to Eq.~(\ref{LET non-SUSY}),  
we can obtain low-energy theorems for two-body 
scattering amplitudes among NG and QNG bosons 
at general points of target spaces. 

At the symmetric point, $v\pri =v$, 
all coefficients become $\1{v^2}$. 
There, all $A^i$ and $A^{N-1}$ fields are NG bosons 
of the symmetry breaking, $O(N)$ to $O(N-1)$, 
and we can verify that 
their scattering amplitudes satisfy 
low-energy theorems from a equation, 
${f_{ij}}^a f_{akl} 
= (\delta_{ik}\delta_{jl}-\delta_{il}\delta_{jk})$.
Their decay constant is $f_{\rm \pi} = v$. 
All $B$ fields correspond to QNG bosons and 
their low-energy theorems coincide with those of NG partners 
as discussed in Sec.~1 and Ref~\cite{HNOO}. 

At the non-symmetric point, 
there appear new features of 
low-energy theorems for NG and QNG bosons.
The unbroken symmetry $O(N-1)$ at symmetric points 
further breaks down to $O(N-2)$, 
and $B^i \;(i=1,\cdots,N-2)$ change to NG bosons for 
the second breaking of $O(N-1)$ to $O(N-2)$. From 
the first equation of (\ref{comp.curvature2}) 
and $R_{B^iB^jB^kB^l}=R_{A^iA^jA^kA^l}$, 
we can verify that low-energy theorems 
for $B^i (i=1,\cdots,N-2)$ coincide with 
those of NG bosons of 
the second symmetry breaking, $O(N-1)$ to $O(N-2)$. 
Low-energy theorems among 
$A^i$ and $A^N$ (at symmetric points) for 
the first breaking of $O(N)$ to $O(N-1)$ 
are distorted there, 
since the field $A^{N-1}$ becomes to play a special role.

\medskip
Before closing this section, 
we give a comment on a relation between 
NG bosons at non-symmetric points in a supersymmetric theory 
and NG bosons in a non-supersymmetric theory. 
In a non-supersymmetric theory with spontaneously broken 
$O(N)$ symmetry to a subgroup $O(N-2)$, 
two sets of linear fields $\vec{\phi}_1$ and $\vec{\phi}_2$, 
belonging to the vector representation, 
should have vacuum expectation values. 
The broken generators are 
\beq
 {\cal G-H}
 = \left(
   \begin{array}{ccc|c|c}
    &                 & &                   &         \\
    & 0               & & f_{\rm \pi}^{(1)} & f_{\rm \pi}^{(2)} \\ 
    &                 & &                   & \\ \hline
    &f_{\rm \pi}^{(1)}& &                   &f_{\rm \pi}^{(3)} \\ \hline
    &f_{\rm \pi}^{(2)}& & f_{\rm \pi}^{(3)} &     
   \end{array}
   \right) , \label{non-susy}
\eeq
where we have expressed 
three $H$-irreducible sectors of broken generators by  
decay constants $f_{\rm \pi}^{(i)}\;(i=1,2,3)$ 
of NG bosons corresponding to these generators.\footnote{
Since there are two vacuum expectation values 
$\vec{v}_1 = \big<\vec{\phi}_1 \big>,
\,\vec{v}_2 = \big<\vec{\phi}_2 \big> \in {\bf N}$ 
to break $O(N)$ to $O(N-2)$, 
these three sectors correspond to three $G$-invariants, 
${\vec{v}_1}^{\,2},\;{\vec{v}_2}^{\,2},\;\vec{v}_1\cdot \vec{v}_2$.
} 
These three free parameters in the non-supersymmetric theory 
are reduced to two parameters, $v$ and $v\pri$, 
to be embedded into a bosonic part of a supersymmetric theory.
This is because there exist $N-2$ pure-type multiplets and 
they relate two decay constants, 
$f_{\rm \pi}^{(1)}$ and $f_{\rm \pi}^{(2)}$, 
in Eq.~(\ref{non-susy}).
This was known at least in pure-realization cases, 
where there exist only pure-type multiplets~\cite{IKK}.

\section{Conclusion and discussion}
If a global symmetry spontaneously 
breaks in supersymmetric theories, 
there appear NG and QNG bosons and 
their fermions superpartners. 
The low-energy effective Lagrangian for these fields 
can be constructed as supersymmetric nonlinear sigma models. 
If symmetry breaking occur by a superpotential of
a fundamental of an effective field theories, 
there must appear at least one QNG boson. 
Hence the target manifold inevitably becomes non-compact. 
As a result, supersymmrtic vacuum alignment occur; 
NG and QNG boson can change with the total number preserved. 
This has been understood by different embedding of 
$\hat H$ into $G^{\bf C}$. 
Low-energy theorems of two-body scattering amplitudes 
for these bosons was known at symmetric points. 

In this paper, we have calculated 
low-energy theorems for NG and QNG bosons 
at general points. 
We have found new features of low-energy theorems. 
In a theory with the supersymmetric vacuum alignment, 
symmetry breaking occurs twice (or more times for other models). 
The low-energy theorems for the first breaking 
at symmetric points have been distorted 
at non-symmetric points; 
on the other hand, low-energy theorems for 
the second breaking coincide with non-supersymmetric cases. 
This is because one (or some for other models) 
NG boson must sit in mixed-type multiplet, 
and play a special role in the sense 
that its partner is QNG boson.  

Although we have illustrated the low-energy theorems 
at non-symmetric points in the $O(N)$-model 
with the simplest (linear) K\"{a}hler potential, 
generalizations to more complicated models are straightforward. 
(The calculation in the most general K\"{a}hler potential 
of the $O(N)$-model is discussed in Appendix B.) 
Consider the case that 
there are $n$ $G^{\bf C}$-invariants and $m$ $G$-invariants. 
(In the $O(N)$-model, they are $m=n=1$, 
since there is one $G^{\bf C}$-invariant, $\vec{\phi}^{\,2}$, 
and one $G$-invariant, $|\vec{\phi}|^2$.) 
The low-energy effective K\"{a}hler potential 
can be written as an arbitrary function 
of $m$ $G$-invariants~\cite{Ni}. 
We can count parameters included in low-energy theorems of 
two-body scattering amplitudes. 
Since curvature tensor includes from one to four derivatives 
of the arbitrary function, 
there are certain numbers of the parameters 
concerned with the arbitrary function. 
Therefore the low-energy theorems include these parameters. 
As seen in this paper, in the case of $O(N)$ model, 
there were six parameters $v,v\pri, f_1, f_2, f_3, f_4$ 
(see Appendix B).

Although we have investigated two-body scattering amplitudes, 
a generalization to many-body scattering amplitudes 
can be calculated by using 
the K\"{a}hler normal coordinate 
to the desired order~\cite{HN2}.  
An interaction Lagrangian can be written by  
the curvature tensor, 
covariant derivatives of the curvature tensor etc. 
If we calculate $n$-body scattering amplitudes, 
it contains from one to $n$ derivatives of the arbitrary function. 

We have investigated only low-energy theorems, 
namely low-energy scattering amplitudes 
at the leading order ${\cal O}(p^2)$.
It is interesting, 
for a development of supersymmetric chiral perturbation theories, 
to investigate higher derivative terms 
such as next-to leading terms 
${\cal O}(p^4)$~\cite{higher}. 
At such order, 
we need a supersymmetric Wess-Zumino-Witten term~\cite{NR}, 
which correctly reproduces anomalies of the global symmetry 
(we did not need it at the lowest order).

\section*{Acknowledgements}
We are grateful to Kazutoshi Ohta and Nobuyoshi Ohta for 
the arguments in the early stage of this work.
The work of M.~N. is supported in part 
by JSPS Research Fellowships.

\begin{appendix}
\section{K\"{a}hler normal coordinate expansion}
In this appendix, 
we show that the K\"{a}hler normal coordinate can be 
always obtained from general coordinates by 
a holomorphic coordinate transformation 
preserving the curvature tensor up to a constant order.
The systematic method to obtain the K\"{a}hler normal coordinate 
to an arbitrary order is discussed in Ref.~\cite{HN2}. 

Let $\{ z^i,z^{*i} \}$ are the general coordinate.
We expand the K\"{a}hler potential by the Taylor expansion as  
\beq
  K(z,z^*)
&=&  K|_0 + F(z) + F^*(z^*) \non
&& + g_{ij^*}|_0\, z^i z^{*j} 
   + \1{2}\Gamma_{i^*jk}|_0\, z^{*i} z^j z^k
   + \1{2} {\Gamma}_{ij^*k^*}|_0\, z^i z^{*j} z^{*k} \non
&& + \1{4} (R_{ij^*kl^*}
       + g_{mn^*}{{\Gamma}^m}_{ik}{{\Gamma}^{n^*}}_{j^*l^*})|_0 \, 
         z^i z^k z^{*j} z^{*l}\non
&& + \1{6}\del_k {\Gamma}_{l^*ij}|_0\, z^i z^j z^k z^{*l} 
   + \1{6}\del_{k^*} {\Gamma}_{li^*j^*}|_0\, 
          z^{*i} z^{*j} z^{*k} z^l + O(z^5) ,\label{TE of K}
\eeq 
where
\beq
 F(z) = \del_i K| \,z^i 
 + \1{2} \del_i \del_j K| \,z^i z^j + \cdots
\eeq
is holomorphic and can be eliminated by 
a K\"{a}hler transformation.
Here ${\Gamma^i}_{jk}$ is the connection and 
$R_{ij^*kl^*}$ is the curvature tensor of 
the K\"{a}hler manifold~\cite{WB}.
There are many non-covariant coefficients except 
for $g_{ij^*}$ and $R_{i^*jk^*l}$.
To eliminate them note that Eq.~(\ref{TE of K}) can be written as 
\beq
 && K(z,z^*) \non 
&&= K|_0  + F(z) + F^*(z^*) \non
&& +g_{mn^*} (z^m + \1{2} {\Gamma^i}_{jk}|z^j z^k 
            + \1{6}\del_k{\Gamma^m}_{ij}| z^i z^j z^k ) 
          (z^n + \1{2} {\Gamma^i}_{jk}|z^j z^k 
            + \1{6}\del_k{\Gamma^n}_{ij}| z^i z^j z^k )^* \non
&& + \1{4} R_{i^*jk^*l}| z^{*i} z^{*k} z^j z^l + O(z^5) .
\eeq
By a holomorphic coordinate transformation
\beq
  \omega^i = z^i + \1{2} {\Gamma^i}_{jk}|z^j z^k
        + \1{6}\del_l {\Gamma^i}_{jk}| z^j z^k z^l + O(z^4), 
   \label{hol.co.tr.}  
\eeq
it can be rewritten as 
\beq
 K(\omega,\omega^*) = K| + 
 \tilde{F}(\omega) + \tilde{F}^*(\omega^*) 
 +g_{ij^*}| \omega^i \omega^{*j} 
  + \1{4} R_{i^*jk^*l}| {\omega}^{*i} {\omega}^{*k} 
                         {\omega}^j {\omega}^l 
 + O({\omega}^5) ,
\eeq
where $\tilde F(\omega) \defeq F(z(\omega))$. 
The new coordinate $\omega$ is 
the K\"{a}hler normal coordinate to the forth order.
In this coordinate, all coefficients are covariant quantities.

By performing these transformations in the superfield level, 
we obtain a K\"{a}hler normal coordinate expansion 
of the Lagrangian given by~\cite{HN2}   
\beq
 {\cal L}
 &=& g_{ij^*}|_{\ph=0}\; \del_{\mu}\ph^i \del^{\mu}\ph^{*j}
 + i g_{ij^*}|_{\ph=0}\; \psb^j \sigb^{\mu} \del_{\mu}\ps^i  \non
 &+& R_{ij^*kl^*}|_{\ph=0}\; 
 \ph^k \ph^{*l} \del_{\mu}\ph^i \del^{\mu}\ph^{*j}
   + \1{4} R_{ij^*kl^*}|_{\ph=0}\; \ps^i\ps^k \psb^j\psb^l \non
 &+& i R_{ij^*kl^*}|_{\ph=0}\;
     \ph^{*j}\del_{\mu}\ph^i (\psb^l \sigb^{\mu} \ps^k ) .
\eeq
which has been used to calculate the low-energy theorems. 
First two terms are equation terms for bosons and fermions, 
and others can be considered interaction terms.
(Although we concentrate on bosonic amplitudes in this paper, 
we can also obtain 
low-energy theorems including fermion partners 
of NG and QNG bosons by using this expansion.) 

Next let us discuss a relation between 
the curvature tensor in an arbitrary coordinate 
and in the K\"{a}hler normal coordinate. 
Since the Jacobian of (\ref{hol.co.tr.})
\beq
 {J^i}_j = \del \omega^i/\del z^j
 = \delta^i_j + O(z) 
\eeq
is unit matrix up to constant order, 
components of the curvature tensor 
in the new coordinate $\omega$ is  
\beq
{R\pri}_{m^*no^*p} = R_{i^*jk^*l} 
 ({J^i}_m)^* {J^j}_n ({J^k}_o)^* {J^l}_p 
 = R_{i^*jk^*l} + O(\omega) \;.
\eeq
Therefore components of the curvature tensor 
is invariant up to constant order. 
We can use an arbitrary coordinate to 
calculate the curvature tensor 
in low-energy theorems (\ref{LET non-SUSY}), 
although low-energy theorems theirselves 
have been obtained in normal coordinates. 

\section{General K\"{a}hler potential} 
In Sec.~4, we have calculated geometric quantities 
in the case of the simplest K\"{a}hler potential 
$K=f(x)=x$.
In this appendix we calculate them 
in an arbitrary case $K=f(x)$.

We can calculate the geometric quantities of 
the $O(N)$ model by using the formulas obtained 
in Sec.~4.1.
First the metric is (we omit primes except for $\vec{v}'$)
\beq
 &&g_{ij^*} = \del_i \del_{j^*} K
            = G_{RS} E^R_i (E^S_j)^* ,\; \non
 &&G_{RS^*}
 = f\pri(z) (\vecvd {Z_S}\dagg \xi\dagg \xi Z_R \vecv)
  +f\prip(z)(\vecvd {Z_S}\dagg \xi\dagg \xi \vecv)
            (\vecvd \xi\dagg \xi Z_R \vecv),  \non
 && z \defeq \vecvd \xi\dagg \xi \vecv.
\eeq
At the point $\ph = 0$, 
we define derivatives of the arbitrary function by 
\beq
 &&f_1 \defeq f\pri(\vecvd \vecv)  = f\pri({v\pri}^2) ,\hs{10} 
 f_2 \defeq f\prip(\vecvd \vecv) = f\prip({v\pri}^2)  
 \;,\; \cdots  \;.   
\eeq
The auxiliary metric at the point $\ph=0$ is
\beq
 G_{RS^*}|_{\ph=0} = f_1 \big<S\dagg R\big> 
         + f_2 \big<S\dagg\big> \big<R\big> \;,
\eeq
where we have used the notations in Eq.~(\ref{notations}). 
This becomes for the $O(N)$ model
\beq
 G_{RS^*}|_{\ph=0}  
 = \pmatrix {f_1 v^2 \delta_{IJ^*} & 0 \cr
            0 & f_1 {v\pri}^2 + f_2 c^4}\;,\label{aux.metric2}
\eeq
\ from Eq.~(\ref{vevs}). 
The vielbein and the ${\hat H}$-connection 
at the point $\ph=0$ are
\beq
 E^R_i|_{\ph =0} = \delta^R_i ,\hs{10} 
 W^M_i|_{\ph =0} = 0,
\eeq
respectively, and differentiations of the vielbein with respect 
to coordinates are
\beq
 \del_j E^R_i|_{\ph =0} = 0  \;.
\eeq

The curvature tensor (\ref{curvature}) on the point $\ph=0$ 
of an arbitrary symmetric $G^{\bf C}/\hat H$ is given by   
\beq
 && \hs{5} R_{RS^*UV^*}|_{\ph=0}  \non
&&= f_1 \big<S\dagg V\dagg UR\big>  
       +f_2 (\big<V\dagg U\big> \big<S\dagg R\big> 
       + \big<S\dagg U\big> \big<V\dagg R\big> 
       + \big<S\dagg V\dagg\big>\big<UR\big> \non
&&\hs{10} 
        + \big<V\dagg\big>\big<S\dagg UR\big> 
        + \big<U\big>\big<S\dagg V\dagg R\big> 
        + \big<S\dagg V\dagg U\big>\big<R\big> 
        + \big<S\dagg\big>\big<V\dagg UR\big> ) \non
&&+ f_3 (\big<V\dagg\big>\big<U\big>\big<S\dagg R\big> 
        + \big<V\dagg\big>\big<S\dagg U\big>\big<R\big>
        + \big<V\dagg\big>\big<S\dagg\big>\big<UR\big> \non
&& \hs{5}   
        + \big<V\dagg U\big>\big<S\dagg\big>\big<R\big>  
        + \big<U\big>\big<S\dagg V\dagg\big>\big<R\big> 
        + \big<U\big>\big<S\dagg\big>\big<V\dagg R\big> )
   + f_4 \big<V\dagg\big>\big<U\big>\big<S\dagg\big>\big<R\big> \non
&&- G^{XY^*}|_{\ph=0} \non
 &&\times 
   (f_1\big<X\dagg SV\big> + f_2(\big<S\big>\big<X\dagg V\big> 
     + \big<X\dagg S\big>\big<V\big> 
   + \big<X\dagg\big>\big<SV\big>) 
   + f_3\big<S\big>\big<X\dagg\big>\big<V\big> )^* \hs{5} \non
&&\times ( f_1\big<Y\dagg RU\big> 
    + f_2(\big<R\big>\big<Y\dagg U\big> 
    + \big<Y\dagg R\big>\big<U\big> 
    + \big<Y\dagg\big>\big<RU\big>) 
    + f_3\big<R\big>\big<Y\dagg\big>\big<U\big> ) ,\hs{10} 
\eeq
where have defined 
\beq
 R_{ij^*kl^*} = R_{RS^*UV^*} 
 \delta^R_i (\delta^S_j)^* \delta^U_k (\delta^V_l)^*  \;. 
\eeq
In the case of the $O(N)$ model, 
it can be calculated from Eq.~(\ref{vevs}), to yield  
\beq
 && R_{IJ^*KL^*} 
    = v^4 \left[{{f_1}^2 + f_1 f_2 {v\pri}^2 
                \over f_1 {v\pri}^2 + f_2 c^4} \right] 
      \delta_{IK}\delta_{J^*L^*}
    + f_2 v^4 (\delta_{IJ^*}\delta_{KL^*} 
    + \delta_{IL^*}\delta_{KJ^*}) \;,\non
 && R_{N-1,J^*,KL^*} = R_{I,N-1,KL^*} = 0 \;,\non
 && R_{N-1,N-1^*,KL^*} 
    = v^2 \left[{f_1 f_2 {v\pri}^2 + (f_1 f_3 - {f_2}^2)c^4 
                 \over f_1 }\right] \delta_{KL^*} \;, \non
 && R_{N-1,J^*,N-1,L^*} 
   = v^4 \left[{{f_1}^2 + f_1 f_2 {v\pri}^2 
               - (2 {f_2}^2 + f_1 f_3) c^4 
              \over f_1 {v\pri}^2 + f_2 c^4}\right] 
     \delta_{J^*L^*} \;, \non
 && R_{N-1,N-1^*,N-1,L^*} = R_{N-1,N-1^*,K,N-1^*} = 0 \;,\non
 && R_{N-1,N-1^*,N-1,N-1^*} \non 
 && = {1 \over f_1 {v\pri}^2 + f_2 c^4} 
    \Big[ 
    {f_1}^2 v^4 + f_1 f_2 {v\pri}^2 (2{v\pri}^4 + v^4 ) 
   - 2 {f_2}^2 ({v\pri}^8 + v^4 {v\pri}^4 - 2v^8) \non
 &&\hspace{2.5cm}
   + 2 f_1 f_3 c^4 (2 {v\pri}^4 + v^4) 
   + f_1 f_4 c^8 {v\pri}^2 
   + (f_2 f_4 - {f_3}^2 )c^{12} \Big],
\eeq
where $c^4 = {v\pri}^4 - v^4$.
We can show that these results 
coincide with direct calculations after solving 
the constraint $\vec{\phi}^2 = a^2$ 
as $\phi^N = \sqrt{a^2 - \sum_{i=1}^{N-1}(\phi^i)^2}$.  
However the coset formalism presented in this paper can be 
generalized to an arbitrary model straightforwardly. 
  
Before the calculation of real components of 
the curvature tensor, we give comments.
\begin{enumerate}
\item
At the symmetric point, 
${v\pri}^2 = v^2,\; c^2 = \sqrt{v^{\prime\,4} - v^4} = 0$, 
the curvatures can be written as 
\beq
 &&R_{ij^*kl^*} = {1\over 2} {f_\pi}^2 \delta_{ik} \delta_{j^*l^*} 
 + g^2 (\delta_{ik}\delta_{j^*l^*} + \delta_{ij^*}\delta_{kl^*}
           +\delta_{il^*}\delta_{kj^*}) \;, \\
 &&{f_\pi}^2 = 2 f_1 v^2 \;,\hs{10} g^2 = f_2 v^4. 
\eeq
We recovered previous results in Ref.~\cite{HNOO} 
and Eq.~(\ref{curve.Kahler}) for the $O(N)$ model.

\item
For the case of the linear K\"{a}hler potential, 
\beq
 f(x) = x \;\; : \;\;f_1 = 1 \;,\; f_2 = f_3 = \cdots = 0 \;,
\eeq
they reduce to results in Eq.~(\ref{comp.curvature}). 
\end{enumerate}

We can calculate 
real components of the curvature tensor 
which are directly concerned with 
low-energy theorems, given by $(i,j,k,l=1,\cdots,N-2)$ 
\beq
&& R_{A^iA^jA^kA^l} 
  = 2 v^4 \left[{{f_1}^2 - {f_2}^2 c^4  
                \over f_1 {v\pri}^2 + f_2 c^4} \right] 
   (\delta_{ik}\delta_{jl} - \delta_{il}\delta_{jk}),  \non
&& R_{B^iA^jB^kA^l} 
  = - 2 v^4 \left[{{f_1}^2 + 2 f_1 f_2 {v\pri}^2 + {f_2}^2 c^4  
                \over f_1 {v\pri}^2 + f_2 c^4} \right] 
         (\delta_{ik}\delta_{jl} + \delta_{il}\delta_{jk})  
         -4 f_2 v^4 (\delta_{ij}\delta_{kl}),  \non
&& R_{B^iA^jA^kA^l} = R_{A^iA^jB^kA^l} = 0, \non
&& R_{A^{N-1}A^jA^{N-1}A^l}  \non 
&& = {2v^2 \over f_1 (f_1 {v\pri}^2 + f_2 c^4)}  
   \Big[{f_1}^3 v^2 + {f_1}^2 f_2 {v\pri}^2 (v^2 - {v\pri}^2 ) 
   - 2 f_1{f_2}^2 v^2 c^4 + {f_2}^3 c^8, \non
 &&\hspace{3.4cm}
   - {f_1}^2 f_3 (v^2 +{v\pri}^2)c^4 - f_1 f_2 f_3 c^8 \Big] 
      \delta_{jl} ,\non
&& R_{A^{N-1}A^{N-1}A^kA^l} =  0, \non
&& R_{B^{N-1}A^jB^{N-1}A^l} \non 
&& = {2v^2 \over f_1 (f_1 {v\pri}^2 + f_2 c^4)}  
   \Big[-{f_1}^3 v^2 - {f_1}^2 f_2 {v\pri}^2 (v^2 + {v\pri}^2 ) 
    + 2 f_1{f_2}^2 v^2 c^4 + {f_2}^3 c^8 \non
 &&\hspace{3.4cm}
   + {f_1}^2 f_3 (v^2 -{v\pri}^2)c^4 - f_1 f_2 f_3 c^8 \Big] 
   \delta_{jl} ,\non
&& R_{B^{N-1}A^{N-1}B^kA^l} 
    = -4 v^2 \left[{f_1 f_2 {v\pri}^2 + (f_1 f_3 - {f_2}^2)c^4 
                 \over f_1 }\right] \delta_{kl}, \non
&& R_{B^{N-1}A^{N-1}B^{N-1}A^{N-1}} \non
 && = {-4 \over f_1 {v\pri}^2 + f_2 c^4} 
    \Big[ 
    {f_1}^2 v^4 + f_1 f_2 {v\pri}^2 (2{v\pri}^4 + v^4 ) 
   - 2 {f_2}^2 ({v\pri}^8 + v^4 {v\pri}^4 - 2v^8) \non
 &&\hspace{2.5cm}
   + 2 f_1 f_3 c^4 (2 {v\pri}^4 + v^4) 
   + f_1 f_4 c^8 {v\pri}^2 
   + (f_2 f_4 - {f_3}^2 )c^{12} \Big] .
\eeq 
At the symmetric point, these reduce to $(i,j,k,l=1,\cdots,N-1)$ 
\beq
&& R_{A^iA^jA^kA^l} 
  = {f_{\pi}}^2 (\delta_{ik}\delta_{jl} 
                  - \delta_{il}\delta_{jk}),  \non
&& R_{B^iA^jB^kA^l} 
  = - {f_{\pi}}^2 (\delta_{ik}\delta_{jl} + \delta_{il}\delta_{jk})  
    - 4 g^2 (\delta_{ik}\delta_{jl} + \delta_{il}\delta_{jk} 
           + \delta_{ij}\delta_{kl}),  \non
&& R_{B^iA^jA^kA^l} = R_{A^iA^jB^kA^l} = 0  \;, 
\eeq      
which again coincide with those obtained in Ref.~\cite{HNOO} 
and Eq.~(\ref{real-curvatures}) for the $O(N)$ model. 

\end{appendix}


\end{document}